\documentclass[twocolumn,showpacs,preprintnumbers,amsmath,amssymb]{revtex4-1}

\usepackage{graphicx}
\usepackage{dcolumn}
\usepackage{amsmath, bm}
\usepackage{amssymb}

\newcommand{ \sH }{ \mathcal{H} }

\newcommand{ \sD }{ \mathcal{D} }

\newcommand{ \sE }{ \mathbb{E} } 

\newcommand{ \half }{ \frac{1}{2} }

\newcommand{ \sgn }{ \mbox{sgn} }

\newcommand{ \Tr }{ \mbox{Tr} }

\begin{document}

\preprint{APS/123-QED}

\title{Robustness of System-Filter Separation for the Feedback Control of a Quantum Harmonic Oscillator Undergoing Continuous Position Measurement}

\author{S. S. Szigeti$^1$}
\author{S. J. Adlong$^1$}
\author{M. R. Hush$^{1,2}$}
\author{A. R. R. Carvalho$^{1,2}$}
\author{J. J. Hope$^{1}$}
\affiliation{$^{1}$Department of Quantum Science, Research School of Physics and Engineering, The Australian National University, ACT 0200, Australia}
\address{$^{2}$ARC Centre for Quantum Computation and Communication Technology, The Australian National University, ACT 0200, Australia}

\date{\today}

\begin{abstract}
We consider the effects of experimental imperfections on the problem of estimation-based feedback control of a trapped particle under continuous position measurement. These limitations violate the assumption that the estimator (i.e. filter) accurately models the underlying system, thus requiring a separate analysis of the system and filter dynamics. We quantify the parameter regimes for stable cooling, and show that the control scheme is robust to detector inefficiency, time delay, technical noise, and miscalibrated parameters. We apply these results to the specific context of a weakly interacting Bose-Einstein condensate (BEC). Given that this system has previously been shown to be less stable than a feedback-cooled BEC with strong interatomic interactions, this result shows that reasonable experimental imperfections do not limit the feasibility of cooling a BEC by continuous measurement and feedback. 
\end{abstract}

\maketitle

\section{Introduction}

Ultracold atomic gases and Bose-Einstein condensates (BEC) are one of the premier platforms for the investigation of quantum fields \cite{Stoof:2009} and precision inertial measurements \cite{Kasevich:2002}, due to their isolation from the environment and ability to be controlled precisely using optical and magnetic fields.  The practical sensitivity of precision devices is limited by the ability to produce stable, low-linewidth sources \cite{Debs:2011, Szigeti:2012}.  Measurement-based feedback control has shown promise for improving control of quantum systems, although early experiments \cite{Mabuchi:2002a,Doherty:2000,Steck:2004,Smith:2002} and much theoretical work \cite{Belavkin:1983,Wiseman:1993,Ramon-van-Handel:2005,Doherty:1999,Wiseman:2010} on feedback control of quantum systems has been applied to relatively low-dimensional systems. Models of Bayesian feedback control on multimode BEC systems have shown that they can be cooled using feedback control for readily accessible trap parameters \cite{Szigeti:2009,Szigeti:2010,Wilson:2007}. The key assumption in most Bayesian feedback control simulations is that the estimate of the state of the system (called the \textit{filter}) is an accurate representation of the system itself.  This assumption is typically very robust, but obviously it can break down outside various limits. This paper quantifies the parameter regime for which this assumption is valid in the context of feedback control of a BEC whose position is continuously monitored. Our results are also applicable to the cooling of nanomechanical resonators \cite{Hopkins:2003}, the localisation of a particle in a double-well potential \cite{Doherty:2000}, and, most generally, to the linear feedback control of a quantum harmonic oscillator undergoing continuous position measurement.   

The first study of feedback control on ultracold gases showed that it could be used to narrow the linewidth of a continuously pumped single-mode atom laser \cite{Wiseman:2001}. This linewidth is normally limited by phase diffusion caused by the strong interatomic nonlinearities.  Feedback significantly reduces this phase diffusion, although the linewidth still scales with the strength of the nonlinearity.  Unfortunately, semiclassical models later showed that continuous pumping would only produce single-mode operation in the presence of strong nonlinearities \cite{Haine:2002,Haine:2003a}. This suggests that alternative methods of stabilising the spatial degrees of freedom are likely to be very productive in producing highly coherent atom laser output.  

Using feedback to control the spatial degrees of freedom of a trapped atom was examined by Doherty and Jacobs \cite{Doherty:1999}, who considered a continuous position measurement of an atom with harmonic confinement.  By assuming an initial Gaussian state for the system, and applying Linear-Quadratic-Gaussian (LQG) control \cite[Sec. 6.4]{Wiseman:2010}, they were able to calculate the optimal cooling scheme even in the presence of measurement backaction. The evolution using an arbitrary initial state was later examined for a linear model \cite{Wilson:2007}, and nonlinear models of a BEC using phase-contrast imaging, which gives a continuous measurement of the density profile rather than just the position \cite{Szigeti:2009,Szigeti:2010}. Cooling to near the ground state was still possible when using this more disruptive measurement process, and the nonlinearities in fact made the cooling more efficient.

All of these simulations used a Bayesian analysis where the best estimate of the quantum state of the system, called the \textit{filter}, was calculated from the dynamics of the system and the measurement result, and an appropriate control was used. This explicit separation of the system and filter has recently been considered in the context of quantum state estimation for a single qubit \cite{Ralph:2011}, a double quantum dot charge state \cite{Cui:2012} and a BEC in a double well potential \cite{Hiller:2012}. Except in pathological cases, it can be shown that the filter converges to the state of the system conditioned on the measurement, so typically no distinction is made between the filter and system for the purposes of discussing controllability of the system. However, in a situation where feedback cooling is competing with uncharacterised heating processes, the timescale of this convergence is particularly important, as the mismatch between the system and filter will be widened by the heating as it is reduced by the measurement. Although filters are typically robust to corruption of the measurement signal, time delays and even miscalibrations of the system, there are obviously limits beyond which the controlled system will no longer be stable. When the filter is not robust to unmodelled uncertainty, control is still possible (although not guaranteed) using \emph{risk-sensitive filtering} \cite{James:2004, James:2008, Yamamoto:2009}, whereby the filter is modified to increase robustness, but with the trade-off that it no longer minimises the least-squares error. Our results show that such an approach is not required, as standard least-squares filtering is robust to corrupting classical Gaussian noise, mismatch of system and filter parameters and time delay of the control signal. Furthermore, this paper quantifies the limits of the controllability of a BEC, and more generally a quantum oscillator, with respect to these experimental inevitabilities.

The paper is structured as follows. In Sec.~\ref{model} we introduce our model of a system-filter separation for linear feedback control of a single particle in an harmonic trap subjected to a continuous measurement of position. Analytic results for the system stability, average steady-state energy and rate of convergence to steady state are presented in Sec.~\ref{Sec_analytics}. These results are used in Sec.~\ref{ExperimentalImperfections} to quantitatively investigate the effects of experimental imperfections on the effectiveness of the control. In particular, we focus on the effects of corrupting classical Gaussian noise, mismatch of system and filter parameters and time delay of the control signal. Finally, in Sec.~\ref{Sec_scenario} we consider the specific example of a trapped, weakly interacting BEC coupled to a cavity mode, and illustrate that a feedback-cooled BEC is robust to experimental imperfections. Sec.~\ref{Conclusions} presents the combined conclusions of these results.

\section{Model of system-filter separation} \label{model}
Semiclassical simulations showed that atomic nonlinearities improved the efficiency of the cooling \cite{Szigeti:2010}, which means that the linear case is in fact the worst-case scenario.  In this limit, we can model a trapped BEC as a single particle, as discussed in more detail in Sec.~\ref{Sec_scenario}.  Therefore, we consider a quantum harmonic oscillator (mass $m$, angular frequency $\omega_S$) undergoing a continuous position measurement and linear feedback control. The system is  described by the Ito stochastic master equation for the conditional density operator $\hat{\rho}_t$ \cite{Doherty:2012}:
\begin{align}
	d\hat{\rho}_t	&= -i [ \hat{H}_0^\rho + \hat{H}_\text{con}^\rho(t), \hat{\rho}_t] dt + \alpha_S \sD\left[ \hat{x} \right]\hat{\rho}_t dt \notag \\
					&+ \sqrt{\alpha_S \eta_S} \, \sH\left[ \hat{x} \right]\hat{\rho}_t dW_t, \label{SME_system}
\end{align}
where $\hat{H}_0^\rho = \left( \hat{p}^2 + \hat{x}^2 \right)/2$ is the oscillator Hamiltonian, $\hat{H}_\text{con}^\rho(t) = u^\rho(t) \hat{x}$ is the linear feedback control Hamiltonian with control signal $u^\rho(t)$, $\alpha_S$ the measurement strength, $\eta_S$ the detector efficiency, $dW_t$ the Wiener increment which satisfies $dW_t dW_{t'} = \delta(t-t') dt$, and the superoperators are defined as
\begin{align}
	\sD\left[\hat{c}\right] \hat{\rho}	&= \hat{c} \hat{\rho} \hat{c}^\dag - \half\left( \hat{c}^\dag \hat{c} \hat{\rho} + \hat{\rho} \hat{c}^\dag \hat{c} \right) \\
	\sH\left[\hat{c}\right] \hat{\rho}	&= \hat{c} \hat{\rho} + \hat{\rho} \hat{c}^\dag - \Tr \left\{ (\hat{c} + \hat{c}^\dag) \hat{\rho} \right\}\hat{\rho}.
\end{align}
These are the decoherence and innovations superoperators respectively for any arbitrary operator $\hat c$. For convenience we have expressed energy and position in units of $\hbar\omega_S$ and $\sqrt{\hbar/(m\omega_S)}$, respectively. 

\begin{figure}[t!]
\centering
\includegraphics[scale=0.55]{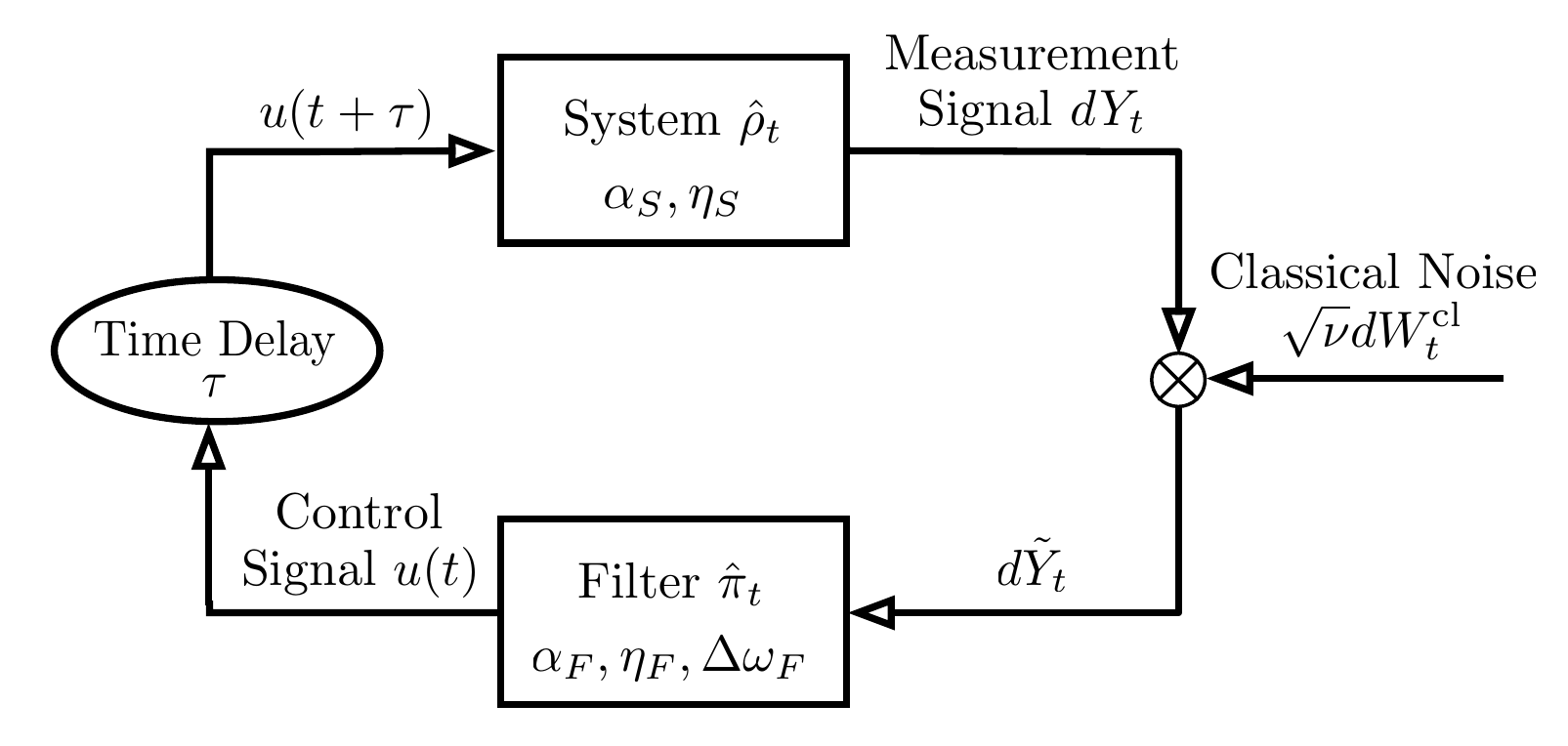}
\caption{\label{system_filter_diagram} Illustration of the feedback control loop under a system-filter separation. A position measurement of the system, $\hat{\rho}_t$, gives a measurement signal $dY_t$. This signal is first corrupted by some classical Gaussian noise $\sqrt{\nu} dW_t^\text{cl}$. The resultant signal, $d\tilde{Y}_t$, is then used to form an estimate of the system, $\hat{\pi}_t$. The control signal $u(t)$, which is a function of the estimate $\hat{\pi}_t$, is fed back into the system after being delayed by some time $\tau$.  }
\end{figure}

For closed-loop feedback control, the control signal must be a function of the \emph{filter} $\hat{\pi}_t$, which is the best-estimate (in the least-squares sense) of the system $\hat{\rho}_t$ conditioned on the information obtained from the continuous position measurement. Ideally, the equation of motion for the filter will be (\ref{SME_system}). However, since filtering requires some \emph{a priori} information about the measurement signal, the system being estimated, and the choice of feedback control, it is possible for the dynamical equation for the filter to differ from that describing the system. We call this a \emph{system-filter separation}. In this paper we consider three distinct experimental imperfections that would result in a system-filter separation:
\begin{enumerate}
	\item \emph{The measurement signal is corrupted by classical noise}. The measurement signal for a continuous position measurement has the form
\begin{equation}
	dY_t = 2\sqrt{\alpha_S \eta_S} \left< \hat{x}\right>_t^\rho + dW_t,
\end{equation}
where $\left< \hat{x} \right>_t^\rho = \Tr \left\{ \hat{x} \hat{\rho}_t\right\}$.  The Gaussian noise on the signal is the irreducible quantum noise from the weak measurement, which is required in order to preserve the commutation relations between the system operators.  It gets relatively smaller as the strength of the measurement $\alpha_S$ is increased, but must always be finite.  However, it is always possible for the position measurement signal to be corrupted by classical noise due to, for example, electronic noise.  If we characterise this noise as Gaussian, then the signal fed into the filter is 
	\begin{equation}
		d\tilde{Y}_t = dY_t + \sqrt{\nu} \, dW_t^\text{cl}, \label{include_cl_noise}
	\end{equation}
where $dW_t^\text{cl}$ is the Wiener increment describing this classical noise and $\nu$ is the strength of the classical noise. 
	\item \emph{Filter and system parameters differ}. The parameters that define the filter are the measurement strength $\alpha_F$, detector efficiency $\eta_F$ and oscillator frequency $\omega_F$. We allow these to differ from the system parameters $\alpha_S, \eta_S$ and $\omega_S$. In harmonic oscillator units, it is more convenient to denote the deviation between the filter and system oscillator frequency with $\Delta \omega_F$, where $\omega_F = \omega_S(1 + \Delta \omega_F)$. 
	\item \emph{Control signal is time-delayed}. In a realistic experiment it will take some finite time to measure the system, construct the estimate, and use this estimate to feed back to the system. This means that the feedback experienced by the system at time $t$ will be based upon the estimate of the system at some prior time $t - \tau$, where $\tau$ is called the time delay. We will assume that the control signal has the form
\begin{equation}
	u^\rho(t) = u(t-\tau) \equiv k \left< \hat{p} \right>_{t-\tau}^\pi, 
\end{equation}
where $\left< \hat{p}\right>_t^\pi = \Tr\left\{\hat{p} \, \hat{\pi}_t \right\}$ is the expectation value of momentum as estimated by the filter, and $k > 0$ is the feedback strength. In contrast, if the experimenter is unaware that there is a time delay, then the filter will model the control signal as 
\begin{equation}
	u^\pi(t) = u(t) = k \left< \hat{p} \right>_t^\pi. 
\end{equation}
We will assume throughout this paper that the feedback strength $k$ can be accurately chosen and implemented by the experimentalist. 
\end{enumerate}
By including these considerations, the equation of motion for the filter is
\begin{align}
	d\hat{\pi}_t	&= -i [ \hat{H}_0^\pi + \hat{H}_\text{con}^\pi, \hat{\pi}_t] dt + \alpha_F \sD\left[ \hat{x} \right]\hat{\pi}_t dt \notag \\
					&+ \sqrt{\alpha_F \eta_F} \, \sH\left[ \hat{x} \right]\hat{\pi}_t (d\tilde{Y}_t - 2\sqrt{\alpha_F \eta_F} \left< \hat{x}\right>_t^\pi), \label{SME_filter}
\end{align}
where $\hat{H}_0^\pi = \left( \hat{p}^2 + (1 + \Delta \omega_F)^2 \hat{x}^2 \right)/2$ and $\hat{H}_\text{con}^\pi(t) = u^\pi(t) \hat{x}$. Note that the innovations term is proportional to the difference between the measurement signal $d\tilde{Y}_t$ and the current best-estimate of the expectation value of position (up to constants). A diagram summarizing the control scheme under a system-filter separation is shown in Fig.~\ref{system_filter_diagram}.

\section{Analytic Results for System-Filter Separation} \label{Sec_analytics}
Since the Hamiltonian for the system contains no terms higher than quadratic order in position and momentum, an initial Gaussian state will remain Gaussian. Furthermore, nonclassical states evolve over a short timescale to Gaussian states due to environmental interactions (such as the measurement process) \cite{Zurek:1993, Garraway:1994, Rigo:1997}. This was verified in the specific context of continuous measurement of atoms by direct integration of the full Wigner function \cite{Wilson:2007}, and the result is that we are free to use the Gaussian approximation.  This approximation allowed Doherty and Jacobs to find the optimal feedback the case when the system and filter are identical \cite{Doherty:1999}, but we will use it to consider the (robust and near optimal) linear damping when the system and filter differ. If $\hat{\rho}_t$ is a Gaussian state then it can be precisely and uniquely represented by the Wigner quasi-probability distribution,
\begin{equation}
	W^\pi(x,p; t) = \frac{\exp\left[ -\frac{1}{2}(\bm{x} - \textbf{x}_t^\rho)^T(\textbf{V}_t^\rho)^{-1}(\bm{x} - \textbf{x}_t^\rho) \right]}{2\pi \sqrt{\det(\textbf{V}_t^\rho)}},
\end{equation}
where $\bm{x}^T = (x,p)$, and
\begin{align}
	\textbf{x}^\rho_t		&= 	\begin{pmatrix}
							\left< \hat{x} \right>_t^\rho \\ \left< \hat{p} \right>_t^\rho
						\end{pmatrix} \\
	\textbf{V}_t^\rho		&=		\begin{pmatrix}
								V_{xx}^\rho(t) 	&	V_{xp}^\rho(t)	\\ 
								V_{xp}^\rho(t)	&	V_{pp}^\rho(t)
							\end{pmatrix}.
\end{align}
Here $\left<\hat{x}\right>^\rho_t$ and $\left<\hat{p}\right>^\rho_t$ are the means, $V_{xx}^\rho = \left<\hat{x}^2\right>^\rho_t - (\left<\hat{x}\right>^\rho_t)^2$ and $V_{pp}^\rho = \left<\hat{p}^2\right>^\rho_t - (\left<\hat{p}\right>^\rho_t)^2$ the variances and $V_{xp}^\rho = \left<\hat{x}\hat{p}+\hat{p}\hat{x}\right>^\rho_t/2 - \left<\hat{x}\right>^\rho_t\left<\hat{p}\right>^\rho_t$ the joint covariance. We refer to the collective of these latter three quantities as the variances. The filter $\hat{\pi}_t$ can also be represented with similarly defined $W^\pi(x,p; t), \textbf{x}_t^\pi$ and $\textbf{V}_t^\pi$.

Under the Gaussian state assumption, it can be shown that Eqs~(\ref{SME_system}) and (\ref{SME_filter}) reduce to matrix differential equations for the means and variances \cite{Doherty:1999, Wilson:2007}:
\begin{subequations}
\label{gaussian_equations}
\begin{align}
	d\textbf{x}_t^\pi	&= \left(\textbf{A}_\pi + \textbf{K} - 4 \eta_F \textbf{V}_t^\pi \textbf{L}_\pi \textbf{L}_\pi^T \right) \textbf{x}_t^\pi dt \notag \\
				     &+ 4\sqrt{\eta_F \eta_S} \, \textbf{V}_t^\pi \textbf{L}_\pi \textbf{L}_\rho^T \textbf{x}_t^\rho dt \notag \\
				     &+ 2 \sqrt{\eta_F} \, \textbf{V}_t^\pi \textbf{L}_\pi \, \left(dW_t + \sqrt{\nu} dW_t^\text{cl}\right) \label{means_filter}\\
	d\textbf{x}_t^\rho	&= (\textbf{A}_\rho\textbf{x}_t^\rho + \textbf{K}\textbf{x}_{t-\tau}^\pi) dt + 2\sqrt{\eta_S} \, \textbf{V}_t^\rho \textbf{L}_\rho \, dW_t \label{means_system}\\
	\dot{\textbf{V}}_t^\pi		&= \textbf{A}_\pi \textbf{V}_t^\pi + \textbf{V}_t^\pi \textbf{A}_\pi^T + \textbf{D}_\pi \notag \\
							&- 4 \eta_F(1+\nu) \textbf{V}_t^\pi \textbf{L}_\pi\textbf{L}^T_\pi \textbf{V}_t^\pi \label{var_pi}\\
	\dot{\textbf{V}}_t^\rho		&= \textbf{A}_\rho \textbf{V}_t^\rho + \textbf{V}_t^\rho \textbf{A}_\rho^T + \textbf{D}_\rho - 4 \eta_S \textbf{V}_t^\rho \textbf{L}_\rho \textbf{L}^T_\rho \textbf{V}_t^\rho, \label{var_rho}
\end{align}
\end{subequations}
where
\begin{subequations}
\begin{equation}
	\bm{\Sigma} = \begin{pmatrix}
		0	&	1 \\
		-1	&	0
	\end{pmatrix},
	\qquad
	\textbf{G} = \begin{pmatrix}
		(1 + \Delta \omega_F)^2	&	0 \\
		0		&	1
	\end{pmatrix},
\end{equation}
\begin{equation}
	\textbf{K} = \begin{pmatrix}
		0	&	0 \\
		0	&	-k
	\end{pmatrix},
	\quad
	\textbf{L}_\pi = 	\begin{pmatrix}
					\sqrt{\alpha}_F	\\	0
				\end{pmatrix},
	\quad
	\textbf{L}_\rho = 	\begin{pmatrix}
					\sqrt{\alpha}_S	\\	0
				\end{pmatrix},
\end{equation}
\end{subequations}
and $\textbf{A}_\pi = \textbf{G}\bm{\Sigma}$, $\textbf{A}_\rho = \bm{\Sigma}$, $\textbf{D}_\pi = \bm{\Sigma} [\textbf{L}_\pi \textbf{L}_\pi^T] \bm{\Sigma}^T$ and $\textbf{D}_\rho = \bm{\Sigma} [\textbf{L}_\rho \textbf{L}_\rho^T] \bm{\Sigma}^T$.

Equations (\ref{var_pi}) and (\ref{var_rho}) are decoupled from each other, the equations for the means, and the feedback. Both are examples of Riccati matrix differential equations, and so are guaranteed to converge to a steady state in the limit $t \to \infty$ \cite{Doherty:1999, Belavkin:1992}:
\begin{subequations}
\label{ss_variances}
\begin{align}
	V_{xx}^\pi(t \to \infty)	&= \frac{1 + \Delta \omega_F}{2\sqrt{2}\alpha_F \eta_F(1+\nu)}\sqrt{\xi_F-1} \\
	V_{pp}^\pi(t \to \infty)	&= \frac{(1+\Delta \omega_F)^3}{2\sqrt{2}\alpha_F \eta_F(1+\nu)}\xi_F \sqrt{\xi_F-1} \\
	V_{xp}^\pi(t \to \infty)	&= \frac{(1+\Delta \omega_F)^2}{4\alpha_F \eta_F (1+\nu)}(\xi_F-1) \\
	 V_{xx}^\rho(t \to \infty)		 &= \frac{1}{2\sqrt{2}\alpha_S \eta_S}\sqrt{\xi_S-1} \\
	V_{pp}^\rho(t \to \infty)		&= \frac{1}{2\sqrt{2}\alpha_S \eta_S}\xi_S \sqrt{\xi_S-1} \\
	V_{xp}^\rho(t \to \infty)		&= \frac{1}{4\alpha_S \eta_S}(\xi_S-1),
\end{align}
\end{subequations}
where for convenience we have defined 
\begin{subequations}
\begin{align}
	\xi_F 	&= \sqrt{1 + \frac{4 \alpha_F^2 \eta_F(1+\nu)}{(1+\Delta \omega_F)^4}} \\
	\xi_S 	&= \sqrt{1 + 4 \alpha_S^2 \eta_S}.
\end{align}
\end{subequations}

In contrast, the equations of motion for the means are delay differential equations, and so have no analytic solution. However, to first order in the time delay, we can we can approximate \cite[pp. 300-301]{Wiseman:2010}
\begin{equation}
	\textbf{x}_{t-\tau}^\pi dt \approx \textbf{x}_t^\pi dt - \tau d\textbf{x}_t^\pi. \label{time_delay_approx}
\end{equation}
This allows Eq.~(\ref{means_system}) to be approximated as a differential equation:
\begin{align}
	d\textbf{x}_t^\rho 	&\approx \textbf{K} \left( (\textbf{I} - \tau\textbf{K}) - \tau \textbf{A}_\pi + 4 \eta_F \tau \textbf{V}_t^\pi \textbf{L}_\pi \textbf{L}_\pi^T\right) \textbf{x}_t^\pi dt \notag \\
						& + \left( \textbf{A}_\rho - 4 \sqrt{\eta_F \eta_S} \,\tau \textbf{K} \textbf{V}_t^\pi \textbf{L}_\pi \textbf{L}_\rho^T\right)\textbf{x}_t^\rho dt \notag \\
						&+2\left( \sqrt{\eta_S}\, \textbf{V}_t^\rho \textbf{L}_\rho - \sqrt{\eta_F} \, \tau \textbf{K} \textbf{V}_t^\pi \textbf{L}_\pi\right)dW_t \notag \\
						&-2\sqrt{\eta_F \nu} \, \tau \textbf{K} \textbf{V}_t^\pi \textbf{L}_\pi dW_t^\text{cl}, \label{means_system_approx}
\end{align}
where $\textbf{I}$ is the $2\times2$ identity matrix. Taking the ensemble average of Eqs~(\ref{means_filter}) and (\ref{means_system_approx}), it can be shown that if a steady state exists, and $k \not = [4\tau\sqrt{\alpha_F \alpha_S \eta_F \eta_S} V_{xp}^\pi(t\to\infty)]^{-1}$, then (see Appendix~\ref{Ap_zero_means})
\begin{equation}
	\sE[\textbf{x}_\infty^\pi] = \sE[\textbf{x}_\infty^\rho] = 0. \label{avg_means_zero}
\end{equation}
However, what we are ultimately concerned with is the average steady-state energy for the system:
\begin{align}
	E_\infty^\rho 	&= \half \sE\left[ \left<\hat{p}^2\right>_\infty^\rho + \left<\hat{x}^2\right>_\infty^\rho \right] \notag \\
				&= \half \sE\left[(\left<\hat{x}\right>_\infty^\rho)^2 + (\left<\hat{p}\right>_\infty^\rho)^2 \right] \notag \\
					&+ \half\left( V_{xx}^\rho(t \to \infty) + V_{pp}^\rho(t \to \infty) \right). \label{steady-state_Energy}
\end{align}
The terms $\sE\left[ (\left<\hat{p}\right>_\infty^\rho)^2\right]$ and $\sE\left[(\left<\hat{x}\right>_\infty^\rho)^2 \right]$ cannot be solved in isolation. However, using $(\ref{means_filter})$ and $(\ref{means_system_approx})$ a closed set of ten coupled differential equations is formed by finding the dynamical equation for the conditional expectation of every pairwise combination of $\left<x\right>_t^\pi$, $\left<p\right>_t^\pi$, $\left<x\right>_t^\rho$ and $\left<p\right>_t^\rho$. These equations can be expressed as the matrix differential equation
\begin{equation}
	\dot{\textbf{v}}_t = \textbf{M}_t\textbf{v}_t + \textbf{b}_t, \label{matrix_eq}
\end{equation} 
where
\begin{widetext}
\begin{align}
	\textbf{v}_t 	&= 	\begin{pmatrix}
					\sE\left[ (\left<x\right>_t^\pi)^2\right] \\ \sE\left[ \left<x\right>_t^\pi \left<p\right>_t^\pi\right] \\ \sE\left[ \left<x\right>_t^\pi \left<x\right>_t^\rho\right] \\ \sE\left[ \left<x\right>_t^\pi \left<p\right>_t^\rho \right] \\ \sE\left[ (\left<p\right>_t^\pi)^2\right] \\ \sE\left[ \left<p\right>_t^\pi \left<x\right>_t^\rho\right] \\ \sE\left[ \left<p\right>_t^\pi \left<p\right>_t^\rho\right] \\ \sE\left[ (\left<x\right>_t^\rho)^2\right] \\ \sE\left[ \left<x\right>_t^\rho \left<p\right>_t^\rho\right] \\ \sE\left[ (\left<p\right>_t^\rho)^2\right]
				\end{pmatrix};
				\qquad
	\textbf{b}_t	= 	\begin{pmatrix}
					\beta_F(1+\nu)(V_{xx}^\pi)^2 \\ \beta_F(1+\nu) V_{xx}^\pi V_{xp}^\pi \\ \beta_{FS} V_{xx}^\pi V_{xx}^\rho \\ \beta_{FS} V_{xx}^\pi V_{xp}^\rho + \beta_F(1+\nu) k \tau V_{xx}^\pi V_{xp}^\pi \\ \beta_F(1+\nu) (V_{xp}^\pi)^2 \\ \beta_{FS} V_{xp}^\pi V_{xx}^\rho \\ \beta_{FS} V_{xp}^\pi V_{xp}^\rho + \beta_F(1+\nu)k \tau (V_{xp}^\pi)^2 \\ \beta_S (V_{xx}^\rho)^2 \\ \beta_S V_{xx}^\rho V_{xp}^\rho + \beta_{FS} k \tau V_{xx}^\rho V_{xp}^\pi \\ \beta_S (V_{xp}^\rho)^2 + 2 \beta_{FS} k \tau V_{xp}^\rho V_{xp}^\pi + \beta_F(1+\nu) k^2 \tau^2 (V_{xp}^\pi)^2
				\end{pmatrix}
\end{align}
\begin{align}
	\textbf{M}_t	&=
	&			\begin{pmatrix}
					-2 \beta_F V_{xx}^\pi 	& 2	& 2 \beta_{FS} V_{xx}^\pi	& 0	& 0	& 0	& 0	& 0	& 0	& 0 \\
					-Q_F	& -(\beta_F V_{xp}^\pi +k)	& \beta_{FS} V_{xp}^\pi	& 0	& 1	& \beta_{FS} V_{xx}^\pi	& 0	& 0	& 0	& 0 \\
					0	& 0	& -\beta_F V_{xx}^\pi & 1	& 0	& 1	& 0	& \beta_{FS} V_{xx}^\pi	& 0	& 0 \\
					-k \tau Q_F	& -k(1+k\tau)	& R_{FS}	& -\beta_F V_{xx}^\pi	& 0	& 0	& 1	& 0	& \beta_F V_{xx}^\pi	& 0 \\
					0	& -2Q_F	& 0	& 0	& -2k	& 2 \beta_{FS} V_{xp}^\pi	& 0	& 0	& 0	& 0 \\
					0	& 0	& -Q_F	& 0	& 0	& -k	& 1	& \beta_{FS} V_{xp}^\pi	& 0	& 0 \\
					0	& -k \tau Q_F	& 0	& -Q_F	& -k(1+k \tau)	& R_{FS}	& -k	& 0	& \beta_{FS} V_{xp}^\pi	& 0 \\
					0	& 0	& 0	& 0	& 0	& 0	& 0	& 0	& 2	& 0 \\
					0	& 0	& -k \tau Q_F	& 0	& 0	& -k(1+k\tau)	& 0	& R_{FS}	& 0	& 1 \\
					0	& 0	& 0	& -2k \tau Q_F	& 0	& 0	& -2k(1+k\tau)	& 0	& 2R_{FS}	& 0
				\end{pmatrix}. \label{M_matrix}
\end{align}
\end{widetext}
For notational compactness we have defined $\beta_F = 4 \alpha_F \eta_F$, $\beta_S = 4 \alpha_S \eta_S$, $\beta_{FS} = 4 \sqrt{\alpha_F \alpha_S \eta_F \eta_S}$ and
\begin{align}
		Q_F 		&= [\beta_F V_{xp}^\pi +(1+\Delta \omega_F)^2] \\
		R_{FS}		&= k \tau \beta_{FS} V_{xp}^\pi-1.
\end{align}

\subsection{Average Steady-State Energy}
In the long term limit, the matrix $\textbf{M}_t$ in Eq.~(\ref{matrix_eq}) goes to a constant matrix $\textbf{M}_\infty$.  Provided $\textbf{M}_\infty$ is invertible, then a unique stationary solution $\textbf{v}_\infty$ exists, given by
\begin{equation}
	\textbf{v}_\infty = -\textbf{M}_\infty^{-1}\textbf{b}_\infty. \label{matrix_ss_soln}
\end{equation}
Note that this is independent of the initial conditions of both the filter and the system. Thus if incorrect initial conditions are input into the filter, then this will not affect the effectiveness of the control in the long time limit. 

In the limit where the system and filter are identical (i.e. $\nu = 0, \tau = 0, \alpha_F = \alpha_S, \eta_F = \eta_S, \Delta \omega_F = 0$), the average steady-state energy is simply that computed for the filter in \cite{Wilson:2007}:
\begin{align}
	E_\infty^0 		&= \alpha_F \eta_F \left\{2 V_{xx}^\pi(t\to \infty) V_{xp}^\pi(t\to\infty) + k [V_{xx}^\pi(t\to\infty)]^2 \right.\notag \\
				& +\left. \frac{1}{2 k \eta_F}\right\} + \half \left[ V_{xx}^\pi(t\to\infty) + V_{pp}^\pi(t\to\infty)\right]. \label{E_ss}
\end{align}
Minimizing Eq.~(\ref{E_ss}) with respect to $k$ gives the optimal feedback strength,
\begin{align}
	k_\text{opt}^\pi 	&= \frac{1}{\sqrt{2 \eta_F} V_{xx}^\pi(t\to\infty)}. \label{k_opt_filter}
\end{align}
Thus an experimenter who believes that the filter represents the underlying system exactly would construct $k_\text{opt}^\pi$ from the filter parameters $\alpha_F$, $\eta_F$ and $\Delta \omega_F$, assume $\nu = 0$, and set $k = k_\text{opt}^\pi$. Explicitly, the experimenter would use the feedback strength
\begin{align}
	k  =  \frac{2\alpha_F \sqrt{\eta_F}}{1 + \Delta \omega_F}\left[\sqrt{1 + \frac{4 \alpha_F^2 \eta_F}{(1+\Delta \omega_F)^4}} -1\right]^{-1/2}. \label{k_opt}
\end{align}
We assume this feedback strength throughout Secs~\ref{ExperimentalImperfections} and \ref{Sec_scenario}. Note, however, that this is only the optimal feedback strength when the system and filter are identical. In general, the optimal feedback strength $k_\text{opt}^\rho$ will be some more complicated function, or cannot be determined analytically. 

\subsection{System Stability}
Even if the stationary solution (\ref{matrix_ss_soln}) exists, there is no guarantee that the system will converge to this steady state. Indeed, we expect there to exist unstable regimes, such as for $k < 0$, which result in gain (and therefore indefinite heating) rather than damping. Heuristically, we expect this to occur when the filter differs from the system to such an extent that the control signal is based on a highly faulty estimate of the system, and is therefore ineffective. We can quantify these regimes of instability by considering a perturbation $\tilde{\textbf{v}}_t = \textbf{v}_t - \textbf{v}_\infty$. Once the variances have attained steady state (which they are guaranteed to do), the equation of motion for this perturbation is
\begin{align}
	\dot{\tilde{\textbf{v}}}_t	&= \textbf{M}_\infty \tilde{\textbf{v}}_t. \label{sys_stability}
\end{align}
As outlined in standard stability analysis, the stability of the system of equations (i.e. whether $\tilde{\textbf{v}}_t$ vanishes, grows indefinitely, oscillates, etc.) is determined by the eigenvalues of the matrix $\textbf{M}_\infty$. If the real component of all the eigenvalues are strictly negative, then $\tilde{\textbf{v}}$ goes to zero, and the system is stable. In this case $\textbf{M}_\infty$ is a negative-definite matrix, hence invertible, and so the steady state (\ref{matrix_ss_soln}) is guaranteed to exist. However, if the real component of at least one eigenvalue of $\textbf{M}_\infty$ is positive, then $\tilde{\textbf{v}}_t$ will grow, and the system will never be controlled to a steady state. 

\subsection{Rate of Convergence to Steady State}
Our control goal is not only to cool the oscillator as close to the ground state as possible, but also as fast as possible. The time it takes for the system energy to converge to steady state depends on (a) the initial conditions, (b) the time it takes for the variances to attain steady state, and (c) the time it takes for $\tilde{\textbf{v}}_t$ to attain steady state. The effect of the initial conditions is a transient that dies out quickly relative to the other timescales, and so its effect can be neglected. The convergence rate of the filter variances to steady state is bounded above by an exponential with decay rate (see Appendix~\ref{Ap_vars})  
\begin{align}
	r_\text{vars}^\pi &= \frac{\sqrt{2}(1+\Delta \omega_F)}{1+\nu}\left(-\sqrt{\xi_F -1} \right.\notag \\
					&\left. + \,\text{Re}\sqrt{\xi_F -1 - \Omega} \right),\label{rate_vars} 
\end{align}
where 
\begin{equation}
	\Omega \equiv 2(1+\nu)\left[1+\nu + (1+\Delta \omega_F)^2(\xi_F-1)\right].
\end{equation}
An identical result holds for $r_\text{vars}^\rho$ with the replacements $\xi_F \to \xi_S$, $\nu \to 0$ and $\Delta \omega_F \to 0$. This is also the rate at which $\textbf{M}_t \to \textbf{M}_\infty$, which occurs before $\tilde{\textbf{v}}_t$ reaches steady state. Finally, the rate at which $\tilde{\textbf{v}}_t$ converges can be estimated by considering the eigenvalues of $\textbf{M}_\infty$. For if $\lambda_i$ and $\textbf{w}_i$ are the eigenvalues and eigenvectors of $\textbf{M}_\infty$, respectively, then 
\begin{equation}
	\tilde{\textbf{v}}_t = \sum_{i=1}^{10} c_i \textbf{w}_i e^{\lambda_i t}
\end{equation}
for constants $c_i$. Hence, for a stable system of equations, in the long time limit (i.e. after all transients have damped out) $|\tilde{\textbf{v}}_t|$ exponentially decays to zero at a rate 
\begin{equation}
	r = -\max_{i}\text{Re}\left\{\lambda_i\right\}. \label{conv_rate}
\end{equation} 
In practice the variances attain steady state well before $|\tilde{\textbf{v}}_t|$. Therefore (\ref{conv_rate}) usually serves as an excellent estimate of the overall timescale. 

\begin{figure}[ht!]
\centering
\includegraphics[scale=0.6]{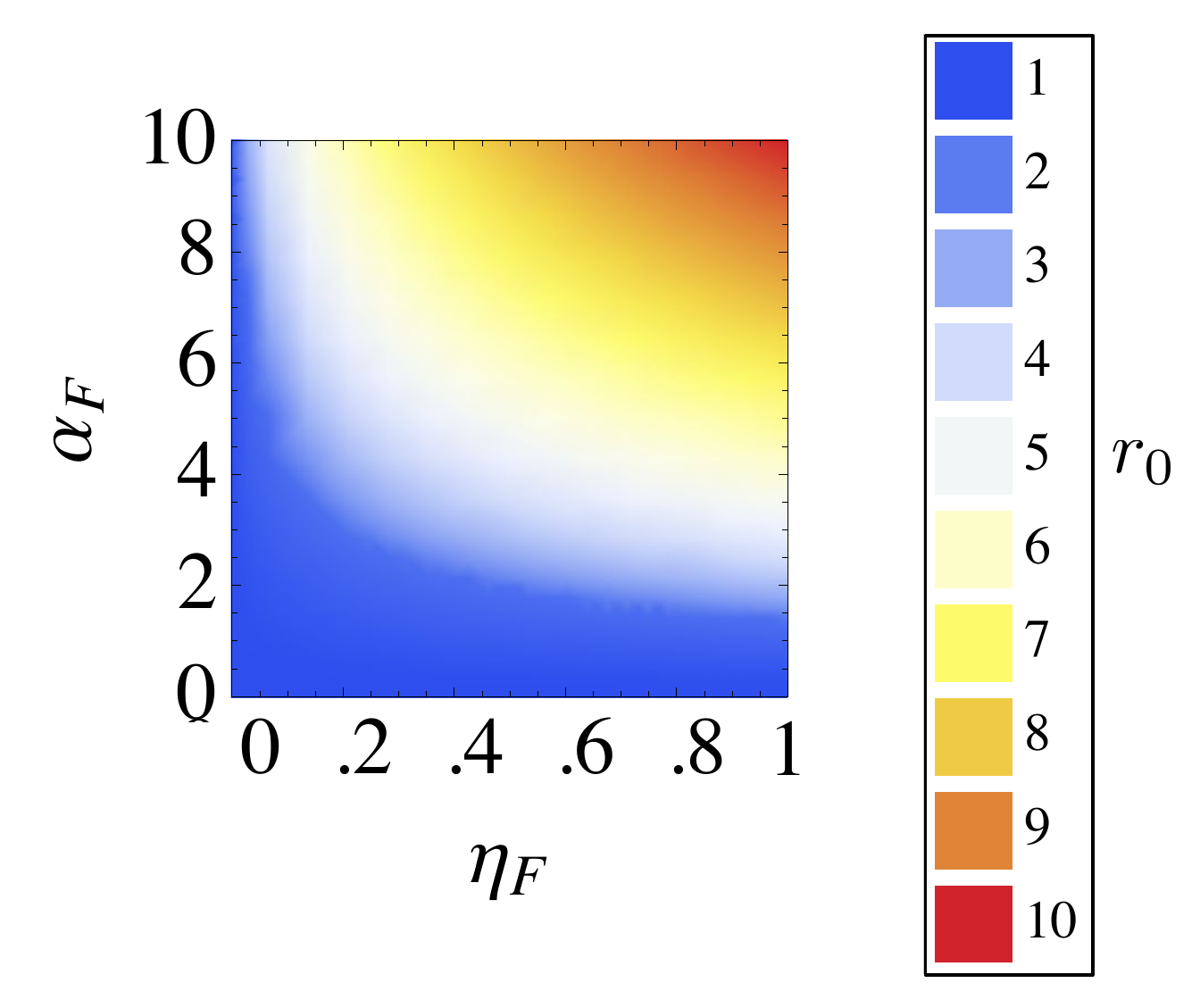}
\caption{\label{plot_r0} Plot of the long time convergence rate $r_0$ as a function of $\alpha_F$ and $\eta_F$, assuming optimal feedback strength $k = k_\text{opt}$.}  
\end{figure}

For the case when the filter and system are identical, the long time convergence rate to steady state can be determined analytically (see Appendix~\ref{Ap_r_0}):
\begin{equation}
	r_0 = \text{Re}\left\{ k + \sqrt{k^2-4}\right\}. \label{r_0}
\end{equation}
Setting the feedback strength to $k_\text{opt}$ [see Eq.~(\ref{k_opt_filter})], a Taylor series expansion of $r_0$ in powers of $\alpha_F \sqrt{\eta_F}$ gives
\begin{equation}
	r_0 = \sqrt{2} + \frac{\alpha_F^2 \eta_F}{\sqrt{2}} + \mathcal{O}\left((\alpha_F \sqrt{\eta_F})^3\right),
\end{equation}
which shows that in the limit of sufficiently small $\alpha_F$ and/or $\eta_F$, the long term convergence rate is independent of the measurement strength and detection efficiency. Outside this regime, as shown in Fig.~\ref{plot_r0}, $r_0$ increases with both increasing $\alpha_F$ (a stronger measurement collapses to steady state faster), and increasing $\eta_F$ (better efficiency improves effectiveness of control), as expected. 

\section{Effects of experimental imperfections}  \label{ExperimentalImperfections}
We now have the tools to consider the effect of classical noise, inaccurate filter parameters and time delay on the efficacy of the feedback control. When judging the effectiveness of the control, we will focus our analysis on the following three criteria:
\begin{enumerate}
	\item Does the system converge to a steady state?
	\item How close to the ground state is the average steady-state energy?
	\item What rate does the system exponentially converge to the steady state?
\end{enumerate}
Doherty and Jacobs \cite{Doherty:1999} and Wilson \emph{et al.} \cite{Wilson:2007} have already addressed these questions in the context where the system and filter are identical. Thus in our analysis, where we consider a separation between the system and filter, results will always be given with respect to this identical case. 

\subsection{Effect of Classical Gaussian Noise} \label{sec_cl_noise}
Let us examine the case where the measurement signal fed into the filter is corrupted by classical Gaussian noise of strength $\nu$ [see Eq.~(\ref{include_cl_noise})], but there is no time delay ($\tau = 0$) and the filter parameters agree with the system ($\alpha \equiv \alpha_F = \alpha_S$, $\eta \equiv \eta_F = \eta_S$ and $\Delta \omega_F = 0$).  Furthermore, the feedback strength $k$ is set by the experimenter under the assumption $\nu = 0$ [see Eq.~(\ref{k_opt})]. In this scenario, the matrix $\textbf{M}_\infty$ is the same as the case when the system and filter are identical, except with the replacement $\eta \to \eta(1+\nu)$ in the steady-state variances for the filter. 

\begin{figure}[t!]
\centering
\includegraphics[scale=0.315]{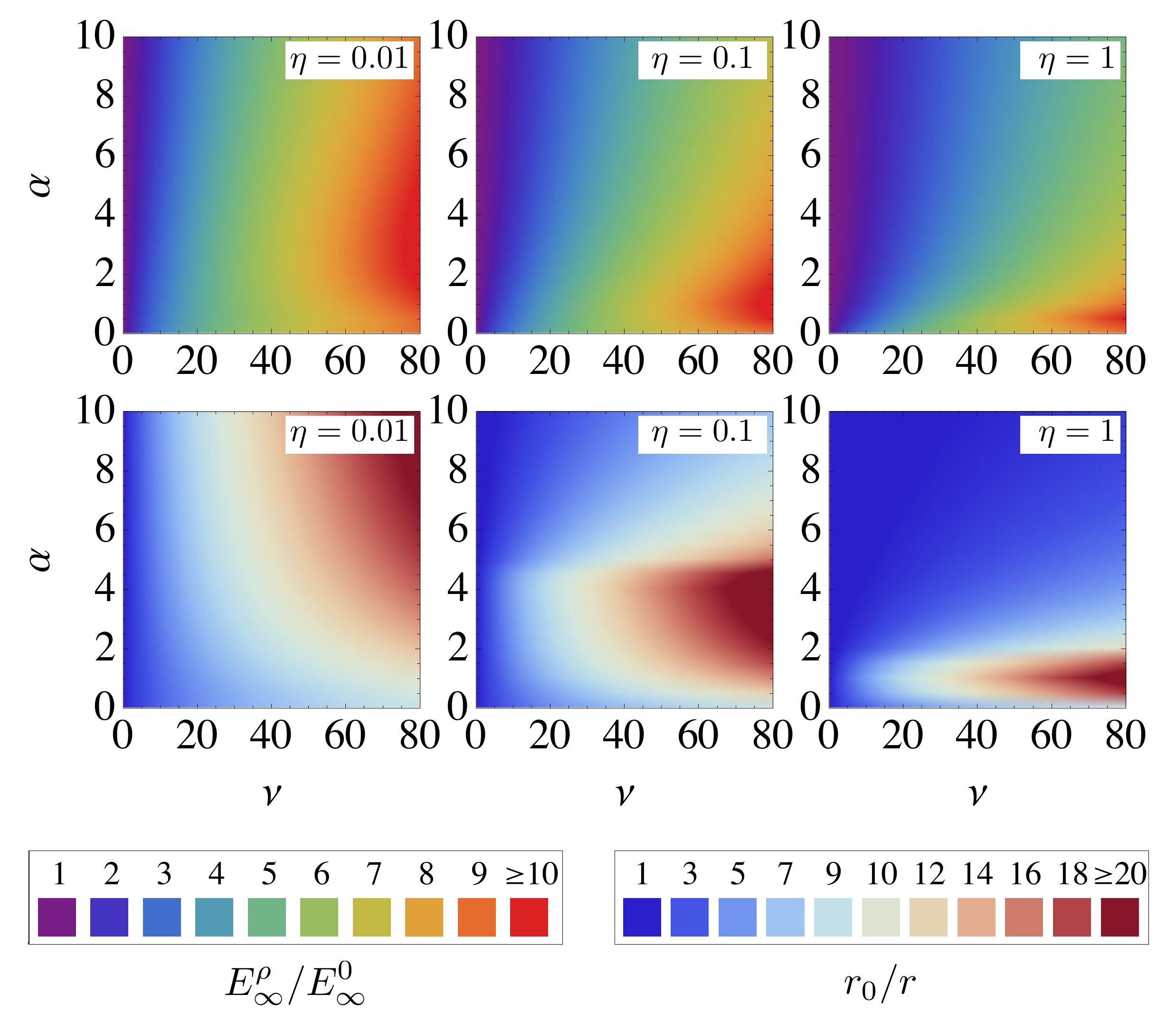}
\caption{\label{noise_plots} Slices of $(\nu, \alpha,\eta)$ parameter space showing the average steady-state energy (top) and long-time convergence rate (bottom). This illustrates the effect of corrupting classical Gaussian noise of strength $\sqrt{\nu}$ on the average steady-state energy and long-time convergence rate, relative to the case when the system and filter are identical [see Eqs~(\ref{E_ss}) and (\ref{r_0})]. Note also that $\tau=0$, $\alpha = \alpha_F = \alpha_S$, $\eta = \eta_F = \eta_S$ and $\Delta \omega_F = 0$.}  
\end{figure}

In the absence of external heating, the corruption of the measurement signal by classical Gaussian noise does not alter the stability, and so the system is guaranteed to converge to a steady state (see Appendix~\ref{Ap_cl_noise_stable}). However, as shown in Fig.~\ref{noise_plots}, the inclusion of this classical noise degrades the effectiveness of the control, both by increasing the average steady-state energy (relative to $E_\infty^0$) and decreasing the rate of convergence to this steady state relative to $r_0$. When the detection efficiency is close to one, the control only loses effectiveness for weaker measurement strengths. However, as $\eta$ is reduced, both $E_\infty^\rho/E_\infty^0$ and $r_0/r$ increase with increasing $\nu$, with little regard for the value of $\alpha$. Note that there are some regimes (e.g. $\alpha = 1$, $\nu = 60$) where decreasing $\eta$ actually decreases $r_0/r$. This does not imply, however, that lower detection efficiencies result in a better control, since both $r$ and $r_0$ decrease with decreasing $\eta$ (\emph{cf.} Fig.~\ref{plot_r0}).   

\subsection{Effect of Imperfect Filter Parameters}
\begin{figure}[ht!]
\centering
\includegraphics[scale=0.315]{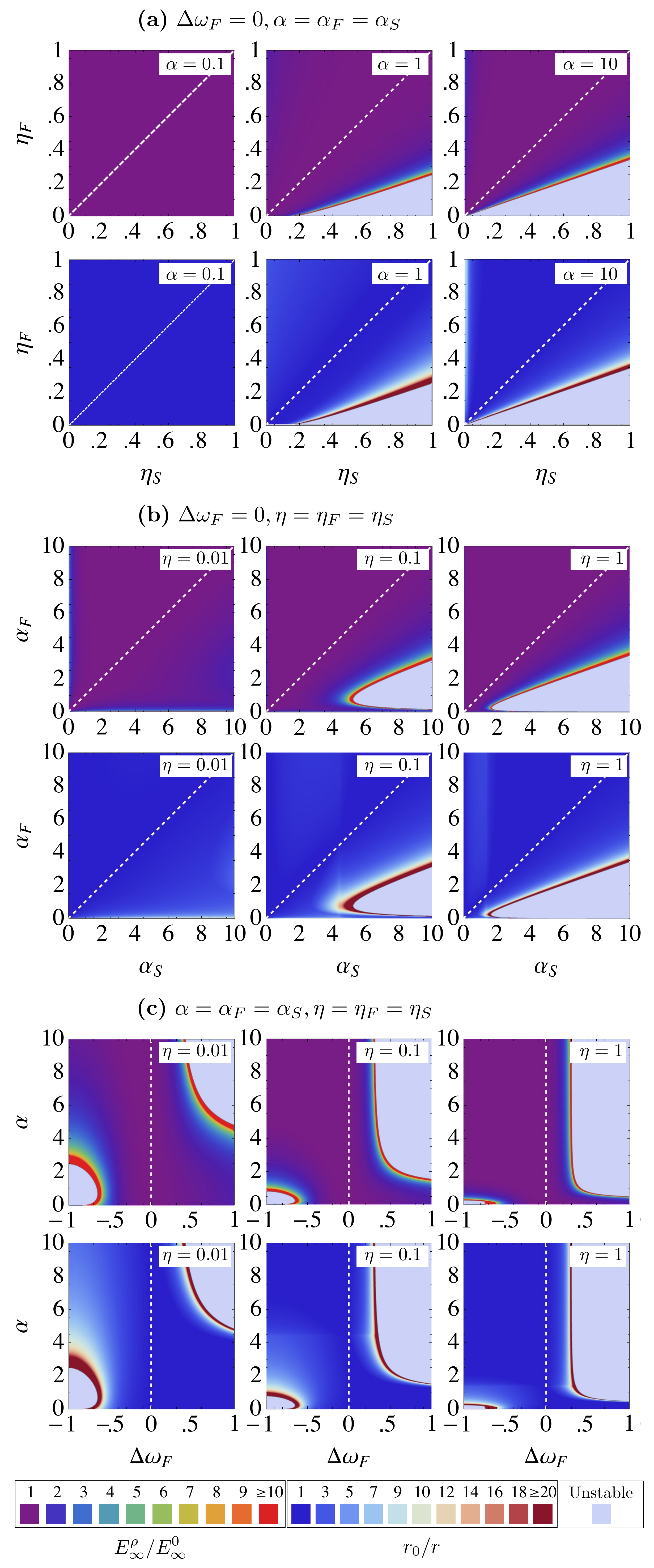}
\caption{\label{plots_imperfect_parameters} Slices in $(\alpha_F,\alpha_S,\eta_F,\eta_S,\Delta \omega_F)$ parameter space showing the average steady-state energy (top) and long-time convergence rate (bottom). This illustrates the effect choosing filter parameters $\alpha_F, \eta_F, \Delta \omega_F$ that differ from the system parameters has on the average steady-state energy and long-time convergence rate, relative to the case when the system and filter are identical [see Eqs~(\ref{E_ss}) and (\ref{r_0})]. The white dashed lines indicate the points where the system and filter are identical. Note also that $\tau = \nu = 0$.}  
\end{figure}

Now we consider the case where there is a difference between the filter and system parameters defining the measurement strength, detection efficiency and oscillator frequency ($\alpha_F \not = \alpha_S, \eta_F \not = \eta_S, \Delta \omega_F \not = 0$), but there is no time delay ($\tau=0$) or classical Gaussian noise ($\nu=0$).  In this scenario, there exist some regimes where the filter's estimate of the system is sufficiently inaccurate that not only is the feedback effectiveness reduced, but is now entirely ineffective. In this regime, the oscillator heats indefinitely, and fails to converge to a steady state. Plots showing regions of stability, average steady-state energies and convergence rates relative to $E_\infty^0$ and $r_0$, respectively, for slices of parameter space $(\alpha_F,\alpha_S,\eta_F,\eta_S,\Delta \omega_F)$ are shown in Fig.~\ref{plots_imperfect_parameters}. 

Before we proceed, let us consider qualitatively what effects a mismatch between the filter and system parameters should have on the effectiveness of the control. For feedback to be effective, the rate at which energy is removed from the oscillator must (eventually) balance the rate at which the measurement backaction causes heating. The rate at which the system heats is fixed by $\alpha_S$. However, the cooling rate is strongly dependent on the filter. All three filter parameters are used to determine the feedback strength (which is optimal if the system and filter agree). Imperfect selection of parameters will result in sub-optimal feedback, which in this context will result in under or over damping of the oscillator. Furthermore, the quality of the estimate $\left<\hat{p}\right>_t^\pi$ depends on $\alpha_F$ and $\eta_F$.  Larger values of these parameters imply that the filter overestimates the momentum diffusion and information rate of the position measurement. 

Let us consider the case $\Delta \omega_F = 0$ (see (a) and (b) in Fig.~\ref{plots_imperfect_parameters}). Firstly, when $\alpha_F > \alpha_S$ and/or $\eta_F > \eta_S$ the system converges to a final steady-state energy approximately equal to the identical system-filter case. Furthermore, the convergence rate $r \approx r_0$. It seems, therefore, that over-estimating the measurement strength and detection efficiency has almost no effect on the effectiveness of the control.  Secondly, if there is a weak measurement strength and/or low detection efficiency, then the control is robust to imperfect guesses of $\alpha_S$ and $\eta_S$. We have seen that filter-based estimation is robust to uncorrelated errors with zero mean, and so it is unsurprising that it is also robust to an incorrectly estimated detection efficiency. The robustness with respect to an incorrectly characterised measurement strength, which causes momentum diffusion, was less clear.

The regime where $\Delta \omega_F \not = 0$ behaves somewhat differently [see Fig.~\ref{plots_imperfect_parameters}(c)].  When the harmonic oscillator frequency is mischaracterised, then even if the filter has a perfect guess of the initial state, then the mean values will diverge from the system. The phase of the feedback will then modulate. Measurement will update the filter to help preserve the phase, but there are obvious regimes of instability if the trap is mischaracterised by a significant fraction.  There is a second effect, whereby the strength of the damping is chosen by the filter, so the feedback response is under- and over-damped on the lower and upper sides of the resonance $\Delta \omega_F = 0$ respectively. The combination of these two effects lead to an asymmetric response with respect to the sign of $\Delta \omega_F$ between the filter and system. Note that there are unstable regions lying on both sides, and a valley of stability near the resonance.

\subsection{Effect of Time Delay}
\begin{figure}[t!]
\centering
\includegraphics[scale=0.33]{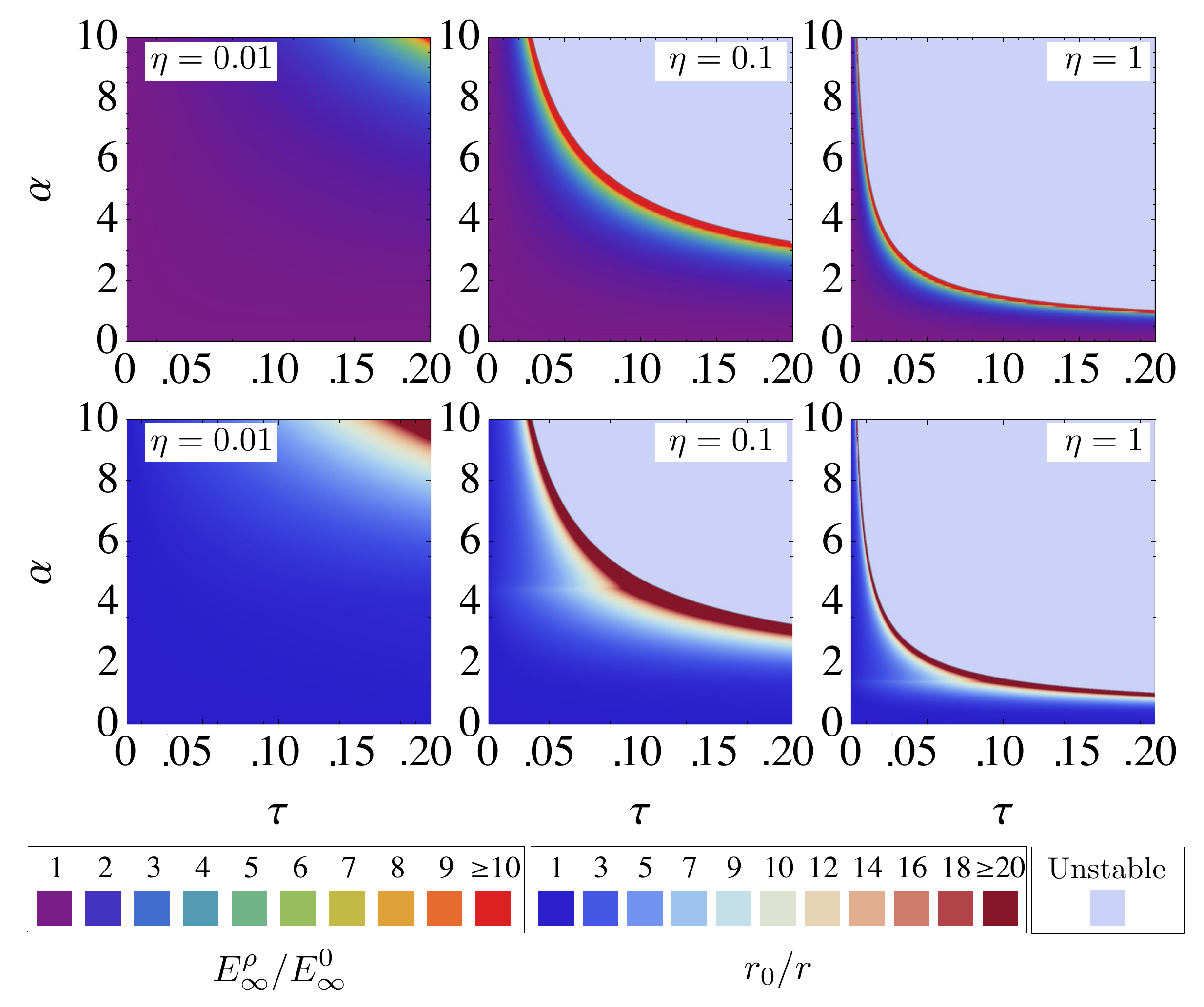}
\caption{\label{plot_1st_order_time_delay} Slices in $(\tau,\alpha,\eta)$ parameter space showing the average steady-state energy (top) and long-time convergence rate (bottom) under short time delay approximation (\ref{time_delay_approx}). This illustrates the effect of a time delay on the average steady-state energy and long-time convergence rate, relative to the case when the system and filter are identical [see Eqs~(\ref{E_ss}) and (\ref{r_0})]. Note also that $\nu = 0$, $\alpha = \alpha_F = \alpha_S$, $\eta = \eta_F=\eta_S$ and $\Delta \omega_F = 0$.}  
\end{figure}

We now consider the case of nonzero time delay $\tau$ when the filter and system parameters agree, and there is no classical noise ($\alpha = \alpha_F = \alpha_S$, $\eta = \eta_F = \eta_S$ and $\nu = \Delta \omega_F = 0$).  By computing the eigenvalues of $\textbf{M}_\infty$, and using Eq.~(\ref{matrix_ss_soln}), we examined the effectiveness of the control (see Fig.~\ref{plot_1st_order_time_delay}). The results show that the control is robust to short time delays, but very quickly loses effectiveness and becomes unstable as $\tau$ increases. However, the boundary of instability is in fact due to a breakdown in the short time-delay approximation (\ref{time_delay_approx}).  We used a numerical solution of the stochastic equations to determine stability for longer delay times.

\begin{figure}[t!]
\centering
\includegraphics[scale=0.315]{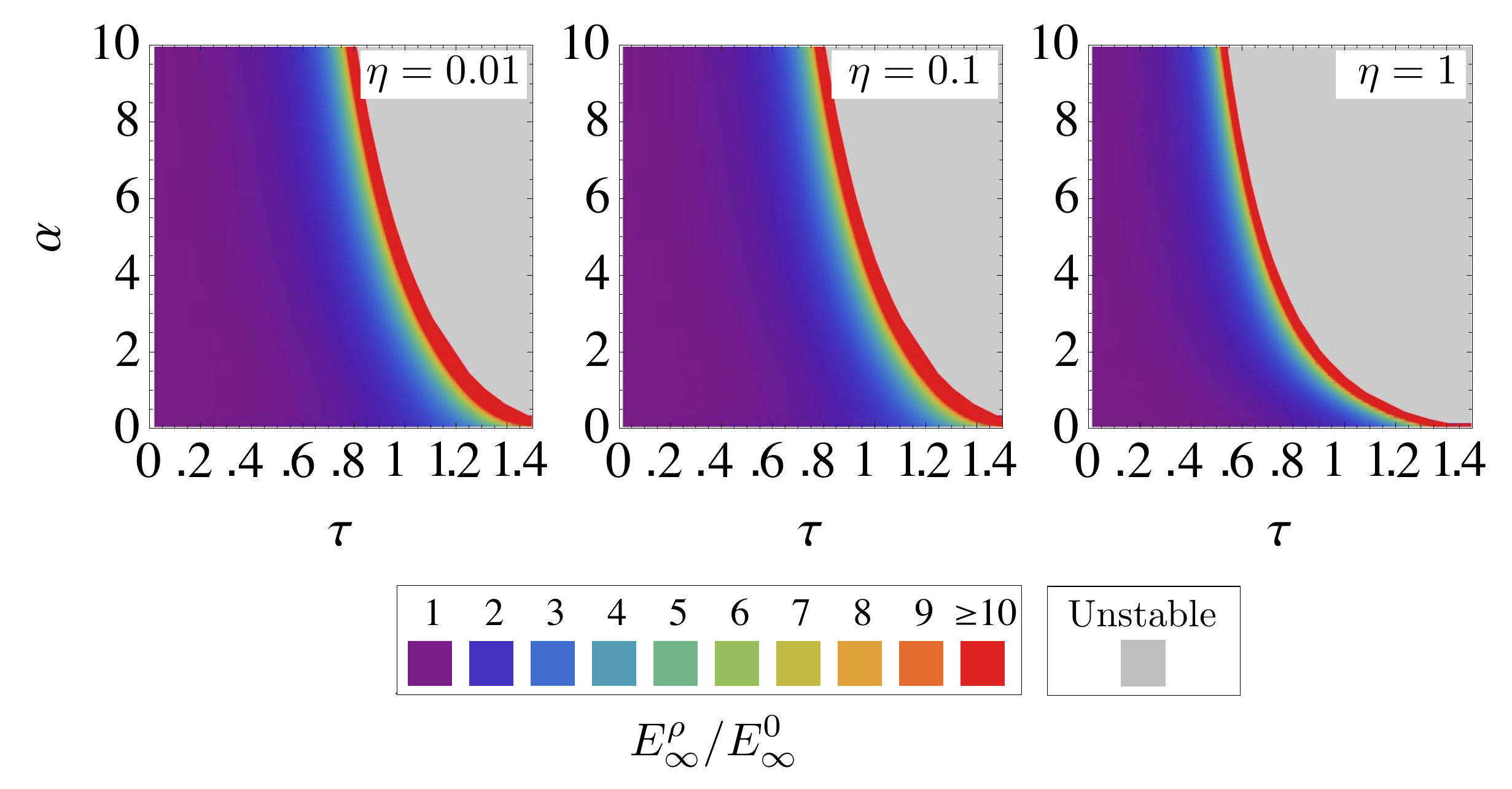}
\caption{\label{numerics} Slices in $(\tau,\alpha,\eta)$ parameter space showing the average steady-state energy, computed via numerical simulation of Eq.~(\ref{means_filter}). Points required averaging over $10^5 - 10^6$ trajectories, and have a standard error no greater than $10\%$. Note also that $\nu = 0$, $\alpha = \alpha_F = \alpha_S$, $\eta = \eta_F=\eta_S$ and $\Delta \omega_F = 0$, and that the $\tau$ axis has a different scale to the plots shown in Fig.~\ref{plot_1st_order_time_delay}.}  
\end{figure}

We computed the average steady-state energy (\ref{steady-state_Energy}) by numerically integrating equations~(\ref{means_filter}) and (\ref{means_system}) with a fixed step 4th order Runge-Kutta algorithm. This was done using the software package \verb+xmds2+ \cite{Dennis:2012}. The results of this numerical analysis are shown in Fig.~\ref{numerics}. It was assumed that the variances had reached the steady-state values (\ref{ss_variances}), and that the means were initially zero.  For the parameters considered, steady state occurred somewhere between $t=10$ and $t=1000$ (where $t$ is in units of $\omega_S^{-1}$). Instability in the system is easily recognised by examining the energy, which when unstable grows exponentially without bound. We categorized a point as unstable if at the end of the interval of integration $E_\infty^\rho/E_\infty^0 > 100$. This heuristic worked well in practice, as most of the unstable points sampled grew to energies $10^6 - 10^8$ times larger than $E_\infty^0$. 

As can be seen from Fig.~\ref{numerics}, the control is stable for larger regimes of parameter space than was predicted under the small time delay approximation (\emph{cf.} Fig.~\ref{plot_1st_order_time_delay}). Instability begins to occur around $\tau \gtrsim 0.6$, which corresponds to the feedback lagging the oscillator by $\gtrsim 35^\circ$. For weaker measurement strengths, the feedback can lag the oscillator a larger amount - for some parameters $\gtrsim 60^\circ$.  Furthermore, the short time-delay approximation (\ref{time_delay_approx}) predicted that as $\tau$ increases there will always be smaller values of $\alpha$ for which the system is stable. The numerical analysis shown in Fig.~\ref{numerics} disagrees, and seems to indicate that after around $\tau \sim 1.4$ feedback will \emph{never} cool the oscillator to a steady state.

\section{Example Scenario: A Feedback-Cooled Bose-Einstein Condensate} \label{Sec_scenario}
\begin{figure}[t!]
\centering
\includegraphics[scale=0.6]{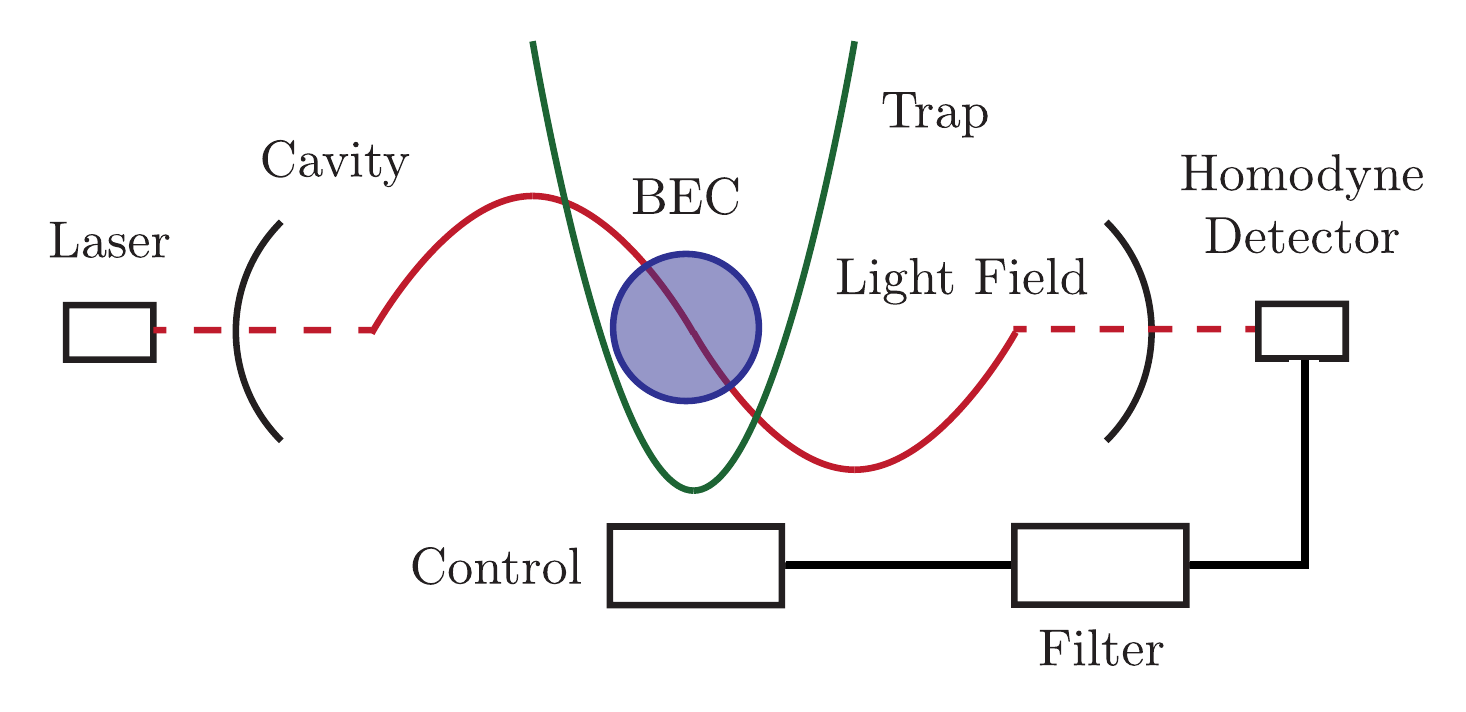}
\caption{\label{BEC_control} Diagram depicting the feedback control of a BEC via a cavity-mediated measurement of the centre-of-mass position.}  
\end{figure}

We now consider a scenario that incorporates all the experimental imperfections examined in Sec.~\ref{ExperimentalImperfections} simultaneously; a feedback-cooled non-interacting BEC. A position measurement could be engineered by placing the condensate in a cavity, probing the cavity with a laser off-resonant with the transition of the BEC, and measuring the output from the cavity (see Fig.~\ref{BEC_control}). More precisely, this gives a measurement of the centre of mass position $\hat{X} = \sum_i \hat{x}_i / N_a$, where $\hat{x}_i$ is the position of the $i$th atom and $N_a$ is the total number of atoms in the condensate. The derivation showing that this physical situation results in a position measurement follows that in Doherty and Jacobs \cite{Doherty:1999} with the replacements $\hat{x} \to \hat{X}$ and $g_0 \to \sqrt{N_a} g_0$ (where $g_0$ is the cavity QED coupling constant). In this case the system state is described by conditional master equation (\ref{SME_system}) with $\hat{x} \to \hat{X}$, $\hat{p} \to \hat{P} = \sum_i \hat{p}_i / N_a$ (for $\hat{p}_i$ the $i$th atom's momentum) and measurement strength
\begin{equation}
	\alpha_S = \frac{4 k_0^2 N_a g_0^4 \bar{n} x_\text{HO}^2}{\omega_S \kappa \Delta^2}, 
\end{equation}
where $k_0 = 2\pi/\lambda$ is the wave vector of the probe laser, $\bar{n}$ the steady state average photon number in the cavity in the absence of the atomic sample, $\kappa$ the cavity linewidth, $\Delta$ the detuning of the probe laser from the cavity resonance and $x_\text{HO} = \sqrt{\hbar/(m\omega_S)}$. Note that this choice of $\alpha_S$ assumes that $\hat{X}$ has been written in units of $x_\text{HO}/\sqrt{N_a}$, which is the natural lengthscale to use in this situation.

We consider a condensate of Rubidium 85 atoms, as this allows us to turn off the interatomic interactions via a Feshbach resonance \cite{Stenger:1999}. A similar situation has been engineered in an optomechanical system, allowing a measurement of the centre of mass position for a (non-condensed) sample of cold atoms \cite{Brahms:2012}. We set the system parameters based loosely upon those used in that experiment: $N_a = 10^4$, $\lambda = 780$ nm, $\omega_S = 2\pi \times 110$ kHz, $g_0 = 2\pi \times 12$ MHz, $\kappa = 2\pi \times 2$ MHz, $\Delta = 2\pi \times 20$ GHz, $\bar{n} = 0.8$ ($\implies \alpha_S \approx 0.1$) and $\eta_S = 0.16$. 

Finally, as a highly conservative upper bound, we will assume that the estimate of the system parameters is incorrect by 100\%: $\alpha_F = 0.05$, $\eta_F = 0.08$ and $\Delta \omega_F = 1$. Furthermore, the measurement fed into the filter has classical Gaussian noise of strength $\nu = 10$, and the control signal is delayed by time $\tau = 0.1$. 

\begin{figure}[ht!]
\centering
\includegraphics[scale=0.5]{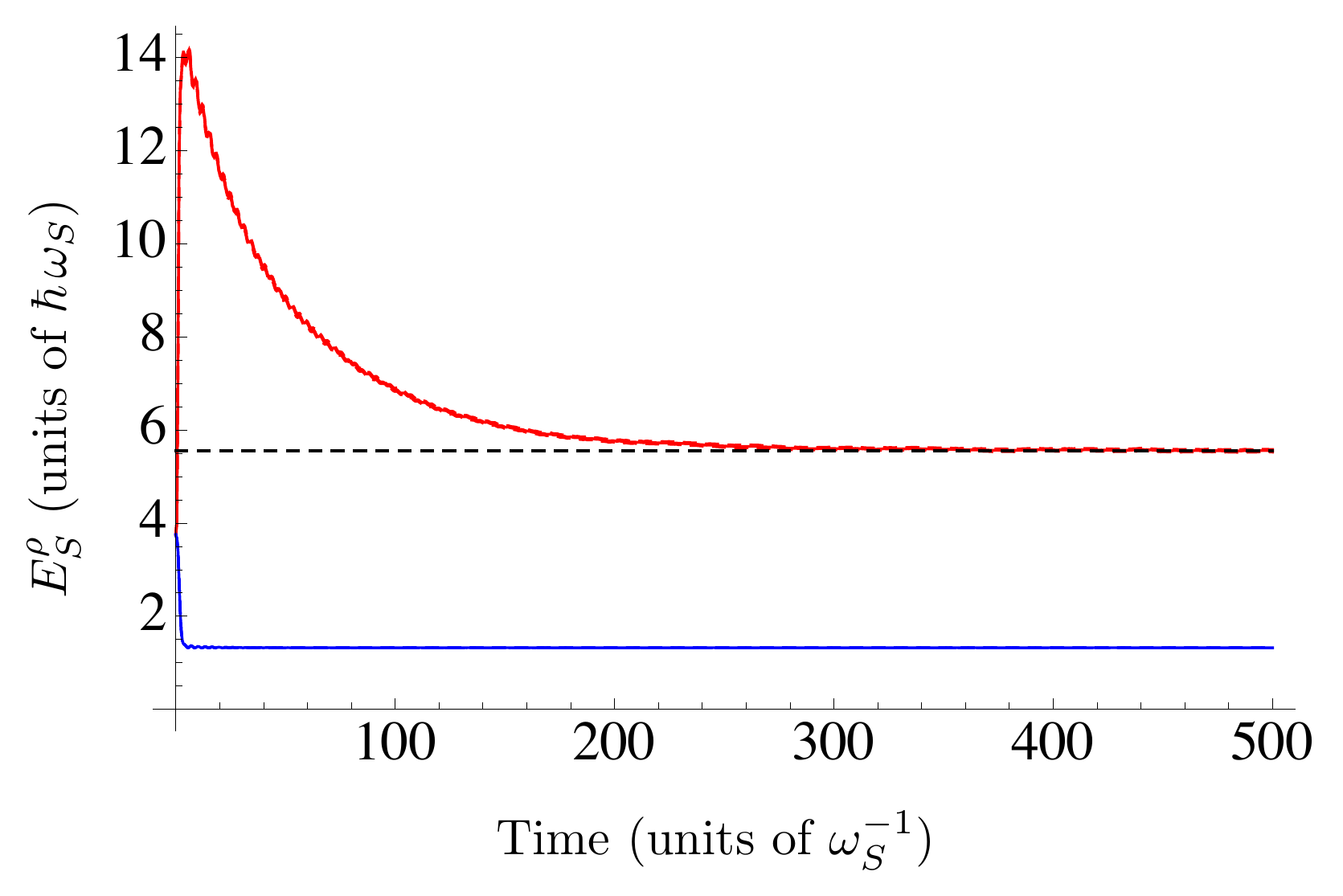}
\caption{\label{scenario_plot} Numerical simulations of the average system energy as a function of time for (blue) identical system and filter ($\alpha_F=\alpha_S = 0.1$, $\eta_S=\eta_F = 0.16$, and $\Delta \omega_F = \nu = \tau = 0$) and (red) a system-filter separation ($\alpha_F = 0.05$, $\alpha_S = 0.1$, $\eta_F = 0.08$, $\eta_S = 0.16$, $\Delta \omega_F = 1$, $\nu = 10$, and $\tau=0.1$). Initial conditions for system and filter were identical, and $\left<\hat{x}\right>^\rho_0 = 2$, $\left<\hat{p}\right>^\rho_0 = 1$, $V_{xx}^\rho(0) = 2$, $V_{xp}^\rho(0) = 0.25$, and $V_{pp}^\rho(0) = 1$. Simulations were over $10^5$ paths, and so the standard error in the means is less than the line thickness on the plots. The black, dashed horizontal line is the average steady-state energy for the system-filter separation (parameters as above) analytically computed using Eq.~(\ref{matrix_ss_soln}).}
\end{figure}

Combining all of these experimental imperfections, numerical simulation of Eqs~(\ref{gaussian_equations}) shows that the control still brings the BEC to a steady state, as is shown by the red curve in Fig.~\ref{scenario_plot}. This of course is in excellent agreement with the analytic results. Compared to the case where the system and filter are identical, the system has an average steady-state energy roughly 4.2 times larger and convergence time $\sim 100$ times longer.

\section{Conclusions} \label{Conclusions}
In this paper we have investigated the linear feedback control of a quantum harmonic oscillator undergoing a continuous position measurement under the more realistic scenario of a filter-system separation (see Fig.~\ref{system_filter_diagram}). In particular, we considered the effectiveness of the control (i.e. whether the oscillator cools to steady state, and if so to what energy and at what rate) when (i) the measurement signal fed into the filter is corrupted by classical Gaussian noise, (ii) the filter parameters governing the measurement strength, detector efficiency, and oscillator trapping frequency differ to those of the system, and (iii) the feedback control signal is time delayed. Although our investigation has found regions where the control is clearly ineffective, these only occur for serious mismatches between the filter and system. This was illustrated by considering the specific example of cooling a BEC. Overall, we conclude that, under the likely experimental imperfections that result in a system-filter separation, the control scheme is robust.

\begin{acknowledgments}
SSS and ARRC thank Dr Moritz Hiller for fruitful discussions. JJH acknowledges the support of the ARC Future Fellowship programme.  ARRC and MRH were supported by the ARC Centre of Excellence for Quantum Computation and Communication Technology (Project number CE110001027).
\end{acknowledgments}

\appendix
\section{Derivation of Eq.~(\ref{avg_means_zero})} \label{Ap_zero_means}
Assuming the existence of a steady state on average, the ensemble average of Eqs~(\ref{means_filter}) and (\ref{means_system_approx}) gives
\begin{subequations}
\label{Ap_A_matrix_eqs}
\begin{align}
	0	&= \textbf{M}_1 \sE\left[ \textbf{x}_\infty^\pi\right] + \textbf{M}_2 \sE\left[ \textbf{x}_\infty^\rho\right] \\
	0	&= \textbf{M}_3 \sE\left[ \textbf{x}_\infty^\pi\right] + \textbf{M}_4 \sE\left[ \textbf{x}_\infty^\rho\right],
\end{align}
\end{subequations}
where
\begin{subequations}
\begin{align}
	\textbf{M}_1	&= \textbf{A}_\pi + \textbf{K} - 4\eta_F \textbf{V}_\infty^\pi \textbf{L}_\pi \textbf{L}_\pi^T \\
	\textbf{M}_2	&= 4 \sqrt{\eta_F \eta_S} \textbf{V}_\infty^\pi \textbf{L}_\pi \textbf{L}_\rho^T \\
	\textbf{M}_3	&= \textbf{K} \left( (\textbf{I} - \tau\textbf{K}) - \tau \textbf{A}_\pi + 4 \eta_F \tau \textbf{V}_t^\pi \textbf{L}_\pi \textbf{L}_\pi^T\right) \\
	\textbf{M}_4	&= \textbf{A}_\rho - 4 \sqrt{\eta_F \eta_S} \,\tau \textbf{K} \textbf{V}_t^\pi \textbf{L}_\pi \textbf{L}_\rho^T.
\end{align}
\end{subequations}
Now
\begin{align}
	\det\left( \textbf{M}_1\right) 	&= \left( 1 + \Delta \omega_F\right)^2 + 4 \alpha_F \eta_F [ k V_{xx}^\pi(t\to \infty) \notag \\
								& + \left( 1 + \Delta \omega_F\right)^2 V_{xp}^\pi(t \to \infty)], \\
	\det\left( \textbf{M}_4 \right)	&= 1 - 4 k \tau \sqrt{\alpha_F \alpha_S \eta_F \eta_S}V_{xp}^\pi(t\to\infty).
\end{align}
which are both guaranteed to be nonzero for $k>0$ and $k \not = [4\tau\sqrt{\alpha_F \alpha_S \eta_F \eta_S} V_{xp}^\pi(t\to\infty)]^{-1}$. We can therefore write
\begin{align}
	\sE\left[ \textbf{x}_\infty^\pi\right]	&= - \textbf{M}_1^{-1}\textbf{M}_2 \sE\left[ \textbf{x}_\infty^\rho\right]
\end{align}
and
\begin{align}
	0	&= \left( \textbf{I}_2 - \textbf{M}_1^{-1}\textbf{M}_2 \textbf{M}_4^{-1} \textbf{M}_3\right)\sE\left[ \textbf{x}_\infty^\pi\right].
\end{align}
It is easily checked that the matrices $\textbf{M}_1^{-1}\textbf{M}_2$ and $\left( \textbf{I}_2 - \textbf{M}_1^{-1}\textbf{M}_2 \textbf{M}_4^{-1} \textbf{M}_3\right)$ are nontrivial, which implies that $\sE\left[ \textbf{x}_\infty^\pi\right] = \sE\left[ \textbf{x}_\infty^\rho\right] = 0$, as required. 

\section{Derivation of Bound on Convergence Rate for Variances} \label{Ap_vars}
Consider the difference between the matrix of variances and its steady state, $\tilde{\textbf{V}}_t^\pi = \textbf{V}_t^\pi - \textbf{V}_\infty^\pi$. Using Eq.~(\ref{var_pi}) and
\begin{align}
	0		&= \textbf{A}_\pi \textbf{V}_\infty^\pi + \textbf{V}_\infty^\pi \textbf{A}_\pi^T + \textbf{D}_\pi \notag \\
			&- 4 \eta_F(1+\nu) \textbf{V}_\infty^\pi \textbf{L}_\pi\textbf{L}^T_\pi \textbf{V}_\infty^\pi, \label{ARE}
\end{align}
it can be shown that
\begin{equation}
	\dot{\tilde{\textbf{V}}}_t^\pi = \tilde{\textbf{A}}_\pi \tilde{\textbf{V}}_t^\pi + \tilde{\textbf{V}}_t^\pi \tilde{\textbf{A}}_\pi^T + 4 \eta_F(1+\nu) \tilde{\textbf{V}}_t^\pi \textbf{L}_\pi\textbf{L}_\pi^T \tilde{\textbf{V}}_t^\pi,
\end{equation}
where $\tilde{\textbf{A}}_\pi = \textbf{A}_\pi - 4 \eta_F(1+\nu) \textbf{V}_\infty^\pi \textbf{L}_\pi \textbf{L}_\pi^T$. This is a Lyapunov differential equation, and has solution
\begin{align}
	\tilde{\textbf{V}}_t^\pi 	&= e^{\tilde{\textbf{A}}_\pi t} \tilde{\textbf{V}}_0^\pi e^{\tilde{\textbf{A}}_\pi^T t} \notag \\
							&+ 4 \eta_F(1+\nu) \int_0^t ds \, e^{\tilde{\textbf{A}}_\pi (t - s)}\tilde{\textbf{V}}_s^\pi \textbf{L}_\pi\textbf{L}_\pi^T \tilde{\textbf{V}}_s^\pi e^{\tilde{\textbf{A}}_\pi^T (t - s)}. \label{vars_soln_appendix}
\end{align} 
The two eigenvalues of $\tilde{\textbf{A}}_\pi$ are
\begin{align}
	\lambda_\pm &= \frac{(1+\Delta \omega_F)}{\sqrt{2}(1+\nu)}\left[-\sqrt{\xi_F -1} \pm \sqrt{\xi_F -1 - \Omega} \right]  	
\end{align}
where 
\begin{equation}
	\Omega \equiv 2(1+\nu)\left[1+\nu + (1+\Delta \omega_F)^2(\xi_F-1)\right].
\end{equation}
The real component of these eigenvalues is always negative, and so the integral on the right hand side of Eq.~(\ref{vars_soln_appendix}) is bounded from above by some constant matrix $\textbf{C}$. Hence
\begin{align}
	\tilde{\textbf{V}}_t^\pi 	&\leq e^{\tilde{\textbf{A}}_\pi t} \left[ \tilde{\textbf{V}}_0^\pi + \textbf{C}\right]e^{\tilde{\textbf{A}}_\pi^T t} \notag \\
							&\sim \exp\left[2 \left(\max_{j=\{+,-\}}\text{Re}\left\{\lambda_j\right\}\right) t\right]\tilde{\textbf{C}} \notag \\
							&=\exp\left(\text{Re}\left\{2\lambda_+\right\} t\right)\tilde{\textbf{C}}, \label{V_bound}
\end{align}
for some constant matrix $\tilde{\textbf{C}}$. This gives the result (\ref{rate_vars}). A similar argument allows one to bound $\tilde{\textbf{V}}_t^\rho$. As an aside, one can see immediately that bound (\ref{V_bound}) is highly insensitive to the initial condition, which can only change the bound by a multiplicative factor. 

\section{Derivation of Convergence Rate $r_0$} \label{Ap_r_0}
Since we are assuming that the filter and system are identical, we drop the superscripts and subscripts $\pi, \rho$. The stochastic differential equation for the means [see Eq.~(\ref{means_system})] is an Ornstein-Uhlenbeck process. The correlation function is \cite[p.~109]{Gardiner:2004} 
\begin{align}
	\sE\left[ \left<\hat{x}\right>_t^2 + \left<\hat{p}\right>_t^2\right]	&= \sE\left[ \textbf{x}_t^T \textbf{x}_t\right] \notag \\
		&= \textbf{x}_0^T e^{-\textbf{A}^T t} e^{-\textbf{A} t}\textbf{x}_0 \notag \\
			&+ 4 \eta \int_0^t ds \, \textbf{L}^T \textbf{V}_s^T e^{-\textbf{A}^T (t-s)} e^{-\textbf{A} (t-s)} \textbf{V}_s \textbf{L} \notag \\
		&\leq \textbf{x}_0^T e^{-\textbf{A}^T t} e^{-\textbf{A} t}\textbf{x}_0 + C,
\end{align}
where $C$ is some positive constant that will have no effect on the convergence time. The first term on the right hand side can be explicitly computed, which gives
\begin{align}
	\sE\left[ \left<\hat{x}_t\right>^2 + \left<\hat{p}_t\right>^2\right]	&\leq \frac{e^{-kt}}{k^2 -4}\left[ -4\left(\left<p\right>_0^2 + k \left<x\right>_0\left<p\right>_0 + \left<x\right>_0^2\right) \right. \notag \\
	& + k\sqrt{k^2-4}\left(\left<x\right>_0^2 - \left<p\right>_0^2\right) \notag \\
	& \times \sinh \left( t\sqrt{k^2-4} \right) \notag \\
	&  + k\left(4 \left<x\right>_0 \left<p\right>_0 + k(\left<x\right>_0^2 + \left<p\right>_0^2)\right) \notag \\
	& \left. \times \cosh \left( t\sqrt{k^2-4} \right) \right] + C.
\end{align}
There are three exponentially decaying rates here: $k$ and $\text{Re}\left\{k \pm \sqrt{k^2-4}\right\}$. The slowest decaying rate is
\begin{equation}
	r_0 = \text{Re}\left\{ k + \sqrt{k^2-4}\right\}.
\end{equation}

\section{Proof of System Stability When Measurement Signal is Corrupted by Classical Gaussian Noise} \label{Ap_cl_noise_stable}
In Sec.~\ref{sec_cl_noise} we claimed that when there is no time delay ($\tau = 0$) and the filter parameters agree with the system ($\alpha \equiv \alpha_F = \alpha_S$, $\eta \equiv \eta_F = \eta_S$ and $\Delta \omega_F = 0$), but $\nu \not = 0$, the system and filter always converge to a steady state. This can be straightfowardly shown by examining the matrix $\textbf{M}$ [see Eq.~(\ref{M_matrix})]. In this regime, the eigenvalues of $\textbf{M}_\infty$ are
\begin{subequations}
\begin{align}
	\lambda_1	&= -k \\
	\lambda_2	&= -k - \sqrt{k^2-4} \\
	\lambda_3	&= -k + \sqrt{k^2-4} \\
	\lambda_4	&= -4 \alpha \eta V_{xx}^\pi \\
	\lambda_5	&= -4 \alpha \eta V_{xx}^\pi -2\sqrt{4\alpha \eta \left( \alpha \eta (V_{xx}^\pi)^2 - V_{xp}^\pi \right)-1} \\
	\lambda_6	&= -4 \alpha \eta V_{xx}^\pi +2\sqrt{4\alpha \eta \left( \alpha \eta (V_{xx}^\pi)^2 - V_{xp}^\pi \right)-1} \\
	\lambda_7	&= \half \left[ -\left( k + 4 \alpha \eta V_{xx}^\pi \right) - \sqrt{\zeta_-} \right] \\
	\lambda_8	&= \half \left[ -\left( k + 4 \alpha \eta V_{xx}^\pi \right) + \sqrt{\zeta_-} \right] \\
	\lambda_9	&= \half \left[ -\left( k + 4 \alpha \eta V_{xx}^\pi \right) - \sqrt{\zeta_+} \right] \\
	\lambda_{10}	&= \half \left[ -\left( k + 4 \alpha \eta V_{xx}^\pi \right) + \sqrt{\zeta_+} \right],
\end{align}
\end{subequations}
where
\begin{align}
	\zeta_\pm	&= \left( k + 4 \alpha \eta V_{xx}^\pi \right)^2 - 8 \left[ 1 + \alpha \eta\left( k V_{xx}^\pi + 2 V_{xp}^\pi \right)\right] \notag \\
			&\pm \sqrt{\left(4 - k^2\right)\left[1 + 4 \alpha \eta \left( V_{xp}^\pi - \alpha \eta (V_{xx}^\pi)^2\right)\right]},
\end{align}
and we have assumed the variances have attained their steady-state values given by Eqs~(\ref{ss_variances}). We will now show that the real component of these eigenvalues is always negative, implying that the system always converges to the steady state.
\begin{itemize}
	\item $\bm{\lambda_1}, \bm{\lambda_2}, \bm{\lambda_4}, \bm{\lambda_5}$: For $\alpha>0, \eta>0$ and $\nu \geq 0$ we have $V_{xx}^\pi >0, V_{xp}^\pi >0$ and $k>0$. Under these conditions, a quick inspection shows the real component of these eigenvalues to be strictly negative.
	\item $\bm{\lambda_3}$: For $k^2 < 4$, $\lambda_3 = -k + i\sqrt{4 - k^2} \implies \text{Re}(\lambda_3) = - k < 0$. For $k^2 > 4$, $\sqrt{k^2 - 4} < k$, and therefore $\text{Re}(\lambda_3) < 0$.
	\item $\bm{\lambda_6}$: Using Eqs~(\ref{ss_variances}) it can be shown that 
	\begin{align}
		\alpha \eta (V_{xx}^\pi)^2 - V_{xp}^\pi	&=-\frac{(1+2\nu)(1+\Delta \omega_F)^2(\xi_F-1)}{8 \alpha \eta(1+\nu)^2} \notag \\
										& < 0.
	\end{align}
	Therefore $4\alpha \eta \left(\alpha \eta (V_{xx}^\pi)^2 - V_{xp}^\pi \right) -1 < 0$, which implies that $\text{Re}(\lambda_6) = -4\alpha \eta V_{xx}^\pi < 0$. \\
	\item $\bm{\lambda_7}, \bm{\lambda_8}, \bm{\lambda_9}, \bm{\lambda_{10}}$: It is enough to show that $\text{Re}\left(\sqrt{\zeta_\pm}\right) < \left( k + 4 \alpha \eta V_{xx}^\pi \right)$. First, note that we can rewrite the expression under the square root in $\zeta_\pm$ as
	\begin{align}
		\kappa 	&\equiv \left(4 - k^2\right)\left[1 + 4 \alpha \eta \left( V_{xp}^\pi - \alpha \eta (V_{xx}^\pi)^2\right)\right]	\notag \\
				&= 4\left[ 1 + \alpha \eta \left( k V_{xx}^\pi + 2 V_{xp}^\pi \right)\right]^2 - \left[ \left( k + 4 \alpha \eta V_{xx}^\pi\right)^2\right. \notag \\
				&\left. + 4 \alpha \eta V_{xp}^\pi \left( k^2 + 4\alpha \eta \left(V_{xp}^\pi + kV_{xx}^\pi \right)\right)\right] \\
				&< 4\left[ 1 + \alpha \eta \left( k V_{xx}^\pi + 2 V_{xp}^\pi \right)\right]^2.  \label{gamma_ineq}
	\end{align}
	If $k^2 < 4$ then $\kappa>0$ and so $\zeta_\pm$ is real. In this case
	\begin{align}
		\sqrt{\zeta_-} 	&< \sqrt{ \left( k + 4 \alpha \eta V_{xx}^\pi \right)^2 - 8 \left[ 1 + \alpha \eta\left( k V_{xx}^\pi + 2 V_{xp}^\pi \right)\right]} \notag \\
					&< k + 4 \alpha \eta V_{xx}^\pi,
		\intertext{and}
		\sqrt{\zeta_+} 	&< \sqrt{ \left( k + 4 \alpha \eta V_{xx}^\pi \right)^2 - 6 \left[ 1 + \alpha \eta\left( k V_{xx}^\pi + 2 V_{xp}^\pi \right)\right]} \notag \\
					&< k + 4 \alpha \eta V_{xx}^\pi,
	\end{align}
	where we have used inequality (\ref{gamma_ineq}) to bound $\sqrt{\zeta_+}$. 
	
	If $k^2 > 4$ then $\kappa<0$, and so $\zeta_\pm$ is complex. In order to compute $\sqrt{\zeta_\pm}$, we note that if $\zeta_\pm = a_\pm + i b_\pm$, then
	\begin{align}
		\text{Re}\left( \sqrt{\zeta_\pm} \right)	&= \frac{1}{\sqrt{2}}\sqrt{\sqrt{a_\pm^2 + b_\pm^2} + a_\pm}, \\
		\text{Im}\left( \sqrt{\zeta_\pm} \right)	&= \frac{\sgn (b_\pm)}{\sqrt{2}}\sqrt{\sqrt{a_\pm^2 + b_\pm^2} - a_\pm}. 
	\end{align}
	For the case under consideration
	\begin{align}
		a_\pm 	&=  \left( k + 4 \alpha \eta V_{xx}^\pi \right)^2 - 8 \left[ 1 + \alpha \eta\left( k V_{xx}^\pi + 2 V_{xp}^\pi \right)\right] \\
		b_\pm	&= \pm \sqrt{-\kappa}.
	\end{align}
	We can write
	\begin{equation}
		 b_\pm^2 = \left(k + 4 \alpha \eta V_{xx}^\pi \right)^2\left[ 1 + 2 \alpha \eta\left( k V_{xx}^\pi + 2 V_{xp}^\pi \right)\right] - \mu,
	\end{equation}
	where
	\begin{align}
		\mu	&= 2\left( k^2-2\right) + 2 \alpha \eta \left[ 4\left( k^2-2\right)V_{xp}^\pi + k\left(k^2+4\right)V_{xx}^\pi\right] \notag \\
				&+4\alpha^2\eta^2V_{xx}^\pi\left[ 8k V_{xp}^\pi + \left( 8+3k^2\right)V_{xx}^\pi\right] \notag \\
				&+ 32 \alpha^3 \eta^3 (V_{xx}^\pi)^2\left( k V_{xx}^\pi + 2 V_{xp}^\pi \right).
	\end{align}
	is guaranteed to be positive since $k^2>4$. Therefore, with some algebraic manipulation,
	\begin{align}
		a_\pm^2 + b_\pm^2	&=\left\{\left( k + 4 \alpha \eta V_{xx}^\pi \right)^2 - 8 \left[ 1 + \alpha \eta\left( k V_{xx}^\pi + 2 V_{xp}^\pi \right)\right]\right\}^2 \notag \\
						&+ \left(k + 4 \alpha \eta V_{xx}^\pi \right)^2\left[ 1 + 2 \alpha \eta\left( k V_{xx}^\pi + 2 V_{xp}^\pi \right)\right] - \mu \notag \\
						&< \left( k + 4 \alpha \eta V_{xx}^\pi \right)^4 + 64\left[ 1 + \alpha \eta\left( k V_{xx}^\pi + 2 V_{xp}^\pi \right)\right]^2 \notag \\
						&-14\left( k + 4 \alpha \eta V_{xx}^\pi \right)^2\left[ 1 + \alpha \eta\left( k V_{xx}^\pi + 2 V_{xp}^\pi \right)\right] \notag \\
						&< \left( k + 4 \alpha \eta V_{xx}^\pi \right)^4 + 64\left[ 1 + \alpha \eta\left( k V_{xx}^\pi + 2 V_{xp}^\pi \right)\right]^2 \notag \\
						&+16\left( k + 4 \alpha \eta V_{xx}^\pi \right)^2\left[ 1 + \alpha \eta\left( k V_{xx}^\pi + 2 V_{xp}^\pi \right)\right] \notag \\
						&= \left\{\left( k + 4 \alpha \eta V_{xx}^\pi \right)^2 + 8 \left[ 1 + \alpha \eta\left( k V_{xx}^\pi + 2 V_{xp}^\pi \right)\right]\right\}^2. \label{ineq_a2+b2}
	\end{align}
	Further algebraic manipulation of inequality (\ref{ineq_a2+b2}) yields
	\begin{equation}
		\sqrt{a_\pm^2 + b_\pm^2} + a_\pm < 2\left( k + 4 \alpha \eta V_{xx}^\pi \right)^2,
	\end{equation}
	and therefore $\text{Re}\left( \sqrt{\zeta_\pm} \right) < \left( k + 4 \alpha \eta V_{xx}^\pi \right)$, as required.
\end{itemize}

\newpage 
\bibliography{pos_paper_stuart_refs}

\begin{thebibliography}{10}

\bibitem{Stoof:2009}
Stoof HTC, Dickerscheid DBM, Gubbels K.
\newblock Ultracold Quantum Fields.
\newblock Springer; 2009.

\bibitem{Kasevich:2002}
Kasevich M.
\newblock Coherence with Atoms.
\newblock Science. 2002;298:1363.

\bibitem{Debs:2011}
Debs JE, Altin PA, Barter TH, D\"oring D, Dennis GR, McDonald G, et~al.
\newblock Cold-atom gravimetry with a {B}ose-{E}instein condensate.
\newblock Phys Rev A. 2011 Sep;84:033610.
\newblock Available from:
  \url{http://link.aps.org/doi/10.1103/PhysRevA.84.033610}.

\bibitem{Szigeti:2012}
Szigeti SS, Debs JE, Hope JJ, Robins NP, Close JD.
\newblock Why momentum width matters for atom interferometry with Bragg pulses.
\newblock New Journal of Physics. 2012;14(2):023009.
\newblock Available from:
  \url{http://stacks.iop.org/1367-2630/14/i=2/a=023009}.

\bibitem{Mabuchi:2002a}
Mabuchi H, Doherty AC.
\newblock Cavity Quantum Electrodynamics: Coherence in Context.
\newblock Science. 2002;298(5597):1372--1377.
\newblock Available from:
  \url{http://www.sciencemag.org/content/298/5597/1372.abstract}.

\bibitem{Doherty:2000}
Doherty AC, Habib S, Jacobs K, Mabuchi H, Tan SM.
\newblock Quantum feedback control and classical control theory.
\newblock Phys Rev A. 2000 Jun;62:012105.
\newblock Available from:
  \url{http://link.aps.org/doi/10.1103/PhysRevA.62.012105}.

\bibitem{Steck:2004}
Steck DA, Jacobs K, Mabuchi H, Bhattacharya T, Habib S.
\newblock Quantum Feedback Control of Atomic Motion in an Optical Cavity.
\newblock Phys Rev Lett. 2004 Jun;92(22):223004.

\bibitem{Smith:2002}
Smith WP, Reiner JE, Orozco LA, Kuhr S, Wiseman HM.
\newblock Capture and Release of a Conditional State of a Cavity QED System by
  Quantum Feedback.
\newblock Phys Rev Lett. 2002 Sep;89:133601.
\newblock Available from:
  \url{http://link.aps.org/doi/10.1103/PhysRevLett.89.133601}.

\bibitem{Belavkin:1983}
Belavkin VP.
\newblock Theory of the Control of Observable Quantum Systems.
\newblock Automatica and Remote Control. 1983;44(2):178--188.

\bibitem{Wiseman:1993}
Wiseman HM, Milburn GJ.
\newblock Quantum theory of optical feedback via homodyne detection.
\newblock Phys Rev Lett. 1993 Feb;70(5):548--551.

\bibitem{Ramon-van-Handel:2005}
Ramon~van Handel JKS, Mabuchi H.
\newblock Modelling and feedback control design for quantum state preparation.
\newblock Journal of Optics B: Quantum and Semiclassical Optics.
  2005;7(10):S179--S197.
\newblock Available from: \url{http://stacks.iop.org/1464-4266/7/S179}.

\bibitem{Doherty:1999}
Doherty AC, Jacobs K.
\newblock Feedback control of quantum systems using continuous state
  estimation.
\newblock Phys Rev A. 1999 Oct;60(4):2700--2711.

\bibitem{Wiseman:2010}
Wiseman HM, Milburn GJ.
\newblock Quantum Measurement and Control.
\newblock Cambridge University Press; 2010.

\bibitem{Szigeti:2009}
Szigeti SS, Hush MR, Carvalho ARR, Hope JJ.
\newblock Continuous measurement feedback control of a Bose-Einstein condensate
  using phase-contrast imaging.
\newblock Physical Review A (Atomic, Molecular, and Optical Physics).
  2009;80(1):013614.
\newblock Available from: \url{http://link.aps.org/abstract/PRA/v80/e013614}.

\bibitem{Szigeti:2010}
Szigeti SS, Hush MR, Carvalho ARR, Hope JJ.
\newblock Feedback control of an interacting Bose-Einstein condensate using
  phase-contrast imaging.
\newblock Phys Rev A. 2010 Oct;82(4):043632.

\bibitem{Wilson:2007}
Wilson SD, Carvalho ARR, Hope JJ, James MR.
\newblock Effects of measurement backaction in the stabilization of a
  Bose-Einstein condensate through feedback.
\newblock Physical Review A (Atomic, Molecular, and Optical Physics).
  2007;76(1):013610.
\newblock Available from: \url{http://link.aps.org/abstract/PRA/v76/e013610}.

\bibitem{Hopkins:2003}
Hopkins A, Jacobs K, Habib S, Schwab K.
\newblock Feedback cooling of a nanomechanical resonator.
\newblock Phys Rev B. 2003 Dec;68:235328.
\newblock Available from:
  \url{http://link.aps.org/doi/10.1103/PhysRevB.68.235328}.

\bibitem{Wiseman:2001}
Wiseman HM, Thomsen LK.
\newblock Reducing the Linewidth of an Atom Laser by Feedback.
\newblock Phys Rev Lett. 2001 Feb;86(7):1143--1147.

\bibitem{Haine:2002}
Haine SA, Hope JJ, Robins NP, Savage CM.
\newblock Stability of Continuously Pumped Atom Lasers.
\newblock Phys Rev Lett. 2002 Apr;88:170403.
\newblock Available from:
  \url{http://link.aps.org/doi/10.1103/PhysRevLett.88.170403}.

\bibitem{Haine:2003a}
Haine SA, Hope JJ.
\newblock Mode selectivity and stability of continuously pumped atom lasers.
\newblock Phys Rev A. 2003 Aug;68:023607.
\newblock Available from:
  \url{http://link.aps.org/doi/10.1103/PhysRevA.68.023607}.

\bibitem{Ralph:2011}
Ralph JF, Jacobs K, Hill CD.
\newblock Frequency tracking and parameter estimation for robust quantum state
  estimation.
\newblock Phys Rev A. 2011 Nov;84:052119.
\newblock Available from:
  \url{http://link.aps.org/doi/10.1103/PhysRevA.84.052119}.

\bibitem{Cui:2012}
Cui W, Lambert N, Ota Y, L{\"u} XY, Xiang ZL, You JQ, et~al.
\newblock Confidence and Backaction in the Quantum Filter Equation.
\newblock arXiv:12073942v2 [quant-ph]. 2012;.

\bibitem{Hiller:2012}
Hiller M, Rehn M, Petruccione F, Buchleitner A, Konrad T.
\newblock Unsharp continuous measurement of a Bose-Einstein condensate: Full
  quantum state estimation and the transition to classicality.
\newblock Phys Rev A. 2012 Sep;86:033624.
\newblock Available from:
  \url{http://link.aps.org/doi/10.1103/PhysRevA.86.033624}.

\bibitem{James:2004}
James MR.
\newblock Risk-sensitive optimal control of quantum systems.
\newblock Phys Rev A. 2004 Mar;69:032108.
\newblock Available from:
  \url{http://link.aps.org/doi/10.1103/PhysRevA.69.032108}.

\bibitem{James:2008}
James MR, Nurdin HI, Petersen IR.
\newblock $H^{\infty }$ Control of Linear Quantum Stochastic Systems.
\newblock Automatic Control, IEEE Transactions on. 2008 sept;53(8):1787 --1803.

\bibitem{Yamamoto:2009}
Yamamoto N, Bouten L.
\newblock Quantum Risk-Sensitive Estimation and Robustness.
\newblock Automatic Control, IEEE Transactions on. 2009 jan;54(1):92 --107.

\bibitem{Doherty:2012}
Doherty AC, Szorkovszky A, Harris GI, Bowen WP.
\newblock The quantum trajectory approach to quantum feedback control of an
  oscillator revisited.
\newblock Philosophical Transactions of the Royal Society A: Mathematical,
  Physical and Engineering Sciences. 2012;370(1979):5338--5353.
\newblock Available from:
  \url{http://rsta.royalsocietypublishing.org/content/370/1979/5338.abstract}.

\bibitem{Zurek:1993}
Zurek WH, Habib S, Paz JP.
\newblock Coherent states via decoherence.
\newblock Phys Rev Lett. 1993 Mar;70:1187--1190.
\newblock Available from:
  \url{http://link.aps.org/doi/10.1103/PhysRevLett.70.1187}.

\bibitem{Garraway:1994}
Garraway BM, Knight PL.
\newblock Evolution of quantum superpositions in open environments: Quantum
  trajectories, jumps, and localization in phase space.
\newblock Phys Rev A. 1994 Sep;50:2548--2563.
\newblock Available from:
  \url{http://link.aps.org/doi/10.1103/PhysRevA.50.2548}.

\bibitem{Rigo:1997}
Rigo M, Alber G, Mota-Furtado F, O'Mahony PF.
\newblock Quantum-state diffusion model and the driven damped nonlinear
  oscillator.
\newblock Phys Rev A. 1997 Mar;55:1665--1673.
\newblock Available from:
  \url{http://link.aps.org/doi/10.1103/PhysRevA.55.1665}.

\bibitem{Belavkin:1992}
Belavkin VP.
\newblock Quantum stochastic calculus and quantum nonlinear filtering.
\newblock Physical Review A. 1992;42(2):171--201.
\newblock Available from:
  \url{http://www.sciencedirect.com/science/article/B6WK9-4CRMB5P-G9/2/c7bf839%
e5a48a6d10547019e47ce8d8d}.

\bibitem{Dennis:2012}
Dennis GR, Hope JJ, Johnsson MT.
\newblock XMDS2: Fast, scalable simulation of coupled stochastic partial
  differential equations.
\newblock Computer Physics Communications. 2013;184(1):201--208.
\newblock Available from:
  \url{http://www.sciencedirect.com/science/article/pii/S0010465512002822}.

\bibitem{Stenger:1999}
Stenger J, Inouye S, Andrews MR, Miesner HJ, Stamper-Kurn DM, Ketterle W.
\newblock Strongly Enhanced Inelastic Collisions in a Bose-Einstein Condensate
  near Feshbach Resonances.
\newblock Phys Rev Lett. 1999 Mar;82:2422--2425.
\newblock Available from:
  \url{http://link.aps.org/doi/10.1103/PhysRevLett.82.2422}.

\bibitem{Brahms:2012}
Brahms N, Botter T, Schreppler S, Brooks DWC, Stamper-Kurn DM.
\newblock Optical Detection of the Quantization of Collective Atomic Motion.
\newblock Phys Rev Lett. 2012 Mar;108:133601.
\newblock Available from:
  \url{http://link.aps.org/doi/10.1103/PhysRevLett.108.133601}.

\bibitem{Gardiner:2004}
Gardiner CW.
\newblock Handbook of Stochastic Methods: For Physics, Chemistry and the
  Natural Sciences.
\newblock 3rd ed. Springer-Verlag; 2004.

\end{thebibliography}

\end{document}